\documentclass[aps,prl,twocolumn,superscriptaddress]{revtex4-1}

\usepackage{graphicx,color}
\usepackage{epsfig}
\usepackage{dcolumn}
\usepackage{amsmath,amssymb,bm}

\usepackage{hyperref,url}
\usepackage[noabbrev]{cleveref}
\usepackage{CJK}
\usepackage{lineno}
\usepackage[english]{babel}
\usepackage[T1]{fontenc}
\usepackage{pifont}
\usepackage{amsfonts}
\usepackage{wasysym}
\usepackage{amsbsy}
\usepackage{psfrag}
\usepackage{color}
\usepackage{multirow}
\usepackage{cases}
\usepackage{stackengine}
\usepackage{float}
\usepackage{blkarray}
\usepackage{cancel}
\usepackage{mathrsfs}
\usepackage{empheq}
\usepackage{lipsum}

\usepackage{titlesec}
\titlespacing*{\section}{0pt}{0.2\baselineskip}{0\baselineskip}

\usepackage{xr}

\usepackage{gensymb}
\usepackage{newtxtext}
\usepackage{newtxmath}
\usepackage[dvipsnames]{xcolor}

\addto\captionsenglish{}

\DeclareMathAlphabet{\textbfsf}{\encodingdefault}{\sfdefault}{bx}{sl}

\hypersetup{
    colorlinks=true,
    linkcolor=blue,
    citecolor=blue,
    urlcolor=blue,
}

\usepackage{fancyhdr}
\begin{document}

\setlength{\belowdisplayskip}{0pt} \setlength{\belowdisplayshortskip}{0pt}
\setlength{\abovedisplayskip}{0pt} \setlength{\abovedisplayshortskip}{0pt}



\title{Harnessing elastic instabilities for enhanced mixing and reaction kinetics in porous media}



\author{Christopher A. Browne}
\affiliation{Department of Chemical and Biological Engineering, Princeton University, Princeton, NJ 08544, USA}

\author{Sujit S. Datta}
\email{ssdatta@princeton.edu}
\affiliation{Department of Chemical and Biological Engineering, Princeton University, Princeton, NJ 08544, USA}


\begin{abstract}
Turbulent flows have been used for millennia to mix solutes~\cite{hoover1950re,ottino1989kinematics}; a familiar example is stirring cream into coffee. However, many energy, environmental, and industrial processes rely on the mixing of solutes in porous media where confinement suppresses inertial turbulence. As a result, mixing is drastically hindered, requiring fluid to permeate long distances for appreciable mixing~\cite{kree2017scalar,heyman2020stretching} and introducing additional steps to drive mixing that can be expensive and environmentally harmful. Here, we demonstrate that this limitation can be overcome just by adding dilute amounts of flexible polymers to the fluid. Flow-driven stretching of the polymers generates an elastic instability (EI), driving turbulent-like chaotic flow fluctuations~\cite{groisman2001,burghelea2004chaotic,burghelea2004mixing,burghelea2007elastic}, despite the pore-scale confinement that prohibits typical inertial turbulence. Using in situ imaging, we show that these fluctuations stretch and fold the fluid within the pores along thin layers (``lamellae'') characterized by sharp solute concentration gradients, driving mixing by diffusion in the pores. This process results in a $3\times$ reduction in the required mixing length, a $6\times$ increase in solute transverse dispersivity, and can be harnessed to increase the rate at which chemical compounds react by $3\times$---enhancements that we rationalize using turbulence-inspired modeling of the underlying transport processes. Our work thereby establishes a simple, robust, versatile, and predictive new way to mix solutes in porous media, with potential applications ranging from large-scale chemical production to environmental remediation.

\end{abstract}

\date{\today}

\maketitle

Being able to efficiently mix solutes in disordered three-dimensional (3D) porous media is critical to a broad range of key energy, environmental, and industrial processes. For example, it controls the rate at which chemical compounds react in porous flow reactors~\cite{maree1985biological,maree1987biological,tennakoon1996electrochemical,ajmera2001microfabricated,noorman2007packed,saito1992high,trapp2006unified,saitoh2008column,meirer2018spatial,kudaibergenov2020flow} used for energy storage~\cite{braff2013membrane} or the production of pharmaceuticals~\cite{saito1992high,trapp2006unified,saitoh2008column}, specialty and ``green'' chemicals~\cite{ajmera2001microfabricated,meirer2018spatial,kudaibergenov2020flow,newman2013role,jensen2017flow,trojanowicz2020flow,vaccaro2017sustainable}, biofuels~\cite{trojanowicz2020flow}, and functional nanomaterials~\cite{marre2010synthesis}. In these cases, the reactive solutes are transported by a fluid of density $\rho$ and dynamic viscosity $\mu$ at a mean velocity $U$ around solid grains of diameter $d_p$, such that the Reynolds number quantifying the ratio of inertial to viscous stresses $\mathrm{Re}\equiv\rho U d_p/\mu\ll1$. Thus, the flow is expected to be laminar and steady over time, and reaction rates are limited by the slow diffusion of reactants with coefficient $\mathcal{D}$ across concentration gradients in the pores. 

Moreover, the P\'eclet number quantifying the ratio of the reactant transport rates by advection to diffusion $\mathrm{Pe}\equiv Ud_p/\mathcal{D}$ is frequently $\gg1$. Under these conditions, ``laminar chaotic advection'' (LCA)~\cite{lee2016passive,song2003experimentalmixing} of reactants through the tortuous pore space gives rise to appreciable mixing downstream only over a large length $l_\mathrm{mix}\sim C d_p$, where the constant $C\gtrsim100$~\cite{perkins1963review,le2013stretching,le2015lamellar,lester2016chaotic,kree2017scalar,heyman2020stretching}. Because this mixing length is primarily set by the fixed geometry of the medium, and cannot be appreciably shortened by changes in the imposed flow speed, it represents a fundamental limit in mixing performance. This limitation is exacerbated for chemical reactions, which proceed over a microscopic kinetic time scale $\tau_{k}$ after mixing, and thus require an additional reactor length $\sim\tau_{k}U$. Consequently, the minimum reactor length to complete the reaction increases with increasing flow speed, creating a fundamental trade-off between increasing throughput and reducing reactor length, which bears a considerable portion of capital costs~\cite{guilbert2021chemical,guilbert2021rectify}. 

A similar challenge arises in many biogeochemical and environmental processes~\cite{smith2008compatibility,kitanidis2012delivery,gomez2015denitrification,datta2009redox,matter2016rapid,szulczewski2012lifetime,stegen2016groundwater,bochet2020iron,borer2018spatial,drescher2013biofilm,anglin2008porous,maree1985biological,maree1987biological,tennakoon1996electrochemical,noorman2007packed,smith2008compatibility,anglin2008porous,datta2009redox,kitanidis2012delivery,szulczewski2012lifetime,drescher2013biofilm,gomez2015denitrification,matter2016rapid,stegen2016groundwater,borer2018spatial,bochet2020iron}, which also rely on the mixing and dispersive spreading of reactive compounds in the subsurface. In these cases, dispersive mixing limitations necessitate the drilling of additional wells, which can be environmentally damaging and prohibitively expensive.

  \begin{figure*}
    \centering
    \includegraphics[width=\textwidth]{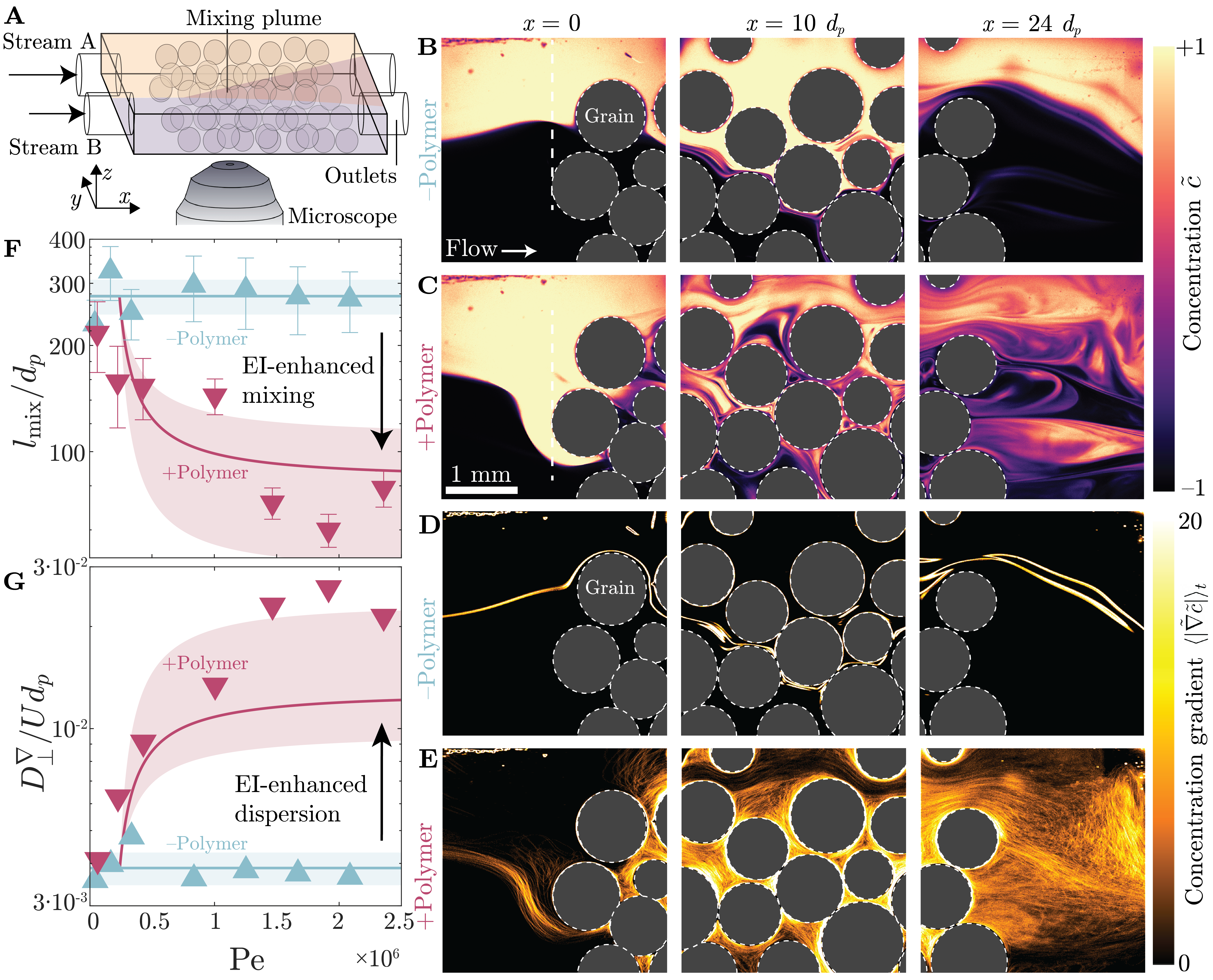}
    \caption{\textbf{In situ imaging reveals that an elastic instability greatly enhances mixing in a porous medium.} \textbf{A} Schematic of the experiment, in which we inject two parallel co-flowing streams of fluid into a model disordered 3D porous medium comprised of sintered glass beads of mean diameter $d_p$. Stream A contains a fluorescent non-reactive dye (yellow in \textbf{B--E}), which acts as a passive solute, while stream B is undyed (black in \textbf{B--E}). We directly visualize the mixing between the streams in the 3D pore space using confocal microscopy, as shown by the optical $x-y$ slices taken at a fixed $z$ position at the inlet ($x=0$), middle ($x=10d_p$), and outlet $x=24d_p$ of the medium in \textbf{B--E}. \textbf{B--C} show the dye concentration, normalized such that $\tilde{c}=+1$ and $-1$ correspond to completely-dyed and -undyed fluid, respectively, and $\tilde{c}=0$ therefore corresponds to completely-mixed fluid, for a mean injection flow speed $U=1.7~\mathrm{mm/s}$. \textbf{D--E} show the corresponding normalized concentration gradient computed in the $x-y$ plane, averaged over the imaging time $t$. In the case of polymer-free solvent, the flow is laminar; as a result, mixing between the streams is minimal (\textbf{B}) and only occurs along a single lamellar surface, characterized by a sharp concentration gradient, along the channel centerline (\textbf{D}). Addition of polymer generates an elastic instability whose chaotic flow fluctuations dramatically improve mixing in the pore space (\textbf{C}), with lamellae of large solute concentration gradient being successively stretched and folded within the individual pores, in addition to across multiple pores (\textbf{E}). As a result, the mixing length $l_\mathrm{mix}$ decreases (\textbf{F}) and the transverse dispersion coefficient $D^{\nabla}_\perp$ increases (\textbf{G}) with increasing P\'eclet number $\mathrm{Pe}$ in the case of a polymeric fluid; error bars on the points represent uncertainty in determining these values from the experiments. The curves show the predictions of our theoretical model, as described further in \emph{Methods}; the shaded region reflects the experimental uncertainty in the values of the model parameters.
    \label{fig:mix}}
\end{figure*}

Here, we introduce a way to overcome these limitations by harnessing an elastic flow instability (EI) generated by polymers added to the fluid. Studies in a range of simplified geometries have shown that the buildup of polymer elastic stresses during transport can generate chaotic flow fields reminiscent of those observed in inertial turbulence, despite the small Reynolds number~\cite{vinogradov1965experimental,larson1990,groisman2000,burghelea2004chaotic,burghelea2004mixing,burghelea2007elastic,scholz2014enhanced,kumar2023lagrangian,kumar2023stress,damseh2010visco,sasmal2023applications,kumar2022transport,walkama2020disorder,haward2021stagnation,browne2021patches,datta2022perspectives}. However, the opacity of more complex 3D porous media precludes visualization of the flow inside the pore space, making similar studies inaccessible. Therefore, it still remains unknown whether---and if so, how---EI can be harnessed to enhances mixing and reaction kinetics in 3D porous media. We overcome this challenge using experiments in model 3D porous media made from sintered random packings of glass beads. In particular, we formulate a dilute polymer solution whose refractive index matches that of the glass, rendering the medium transparent and enabling direct flow visualization in situ using confocal microscopy (detailed in \emph{Methods}) ~\cite{browne2021patches,browne2023homogenizing}.\\

\noindent\textbf{EI greatly enhances solute mixing in a porous medium.} To characterize solute mixing, we inject two parallel co-flowing streams of fluid into the porous medium at the same imposed volumetric flow rate $Q/2$. The mean flow velocity is then $U\equiv Q/(\phi A)$, where $\phi$ and $A$ are the porosity and cross-sectional area of the porous medium, respectively. As shown in Figure~\ref{fig:mix}A, stream A contains a non-reactive dye, which acts as a passive solute, at a concentration $c_A$, while stream B is undyed. Visualizing the dye fluorescence intensity at the interface between them therefore provides a measure of the dye concentration $c$, and therefore the extent of mixing between the streams, as it varies with position $x$ along the length of the medium. To this end, we define a dimensionless concentration $\tilde{c}\left(\mathbf{x},t\right)=\left(c\left(\mathbf{x},t\right)-c_\infty\right)/c_A$, where $\mathbf{x}$ is the $(x,y)$ position vector and $c_\infty\equiv c_A/2$ is the completely-mixed dye concentration, such that $\tilde{c}=1$ and $\tilde{c}=-1$ correspond to the fully dyed and undyed streams and $\tilde{c}=0$ represents complete mixing. Quantifying the increase in the extent of mixing with $x$ then yields a direct measure of the mixing length $l_\mathrm{mix}$~\cite{chertkov2003decay,son1999turbulent,balkovsky1999universal,voth2003mixing,rothstein1999persistent,ran2021bacteria,groisman2001,burghelea2004mixing,burghelea2004chaotic}. Moreover, because mixing is ultimately driven by solute diffusion across short-ranged, sub pore-scale concentration gradients, we use the images to compute $\tilde{\nabla}\tilde{c}=d_p^{-1}\left(\partial_x+\partial_y\right)\tilde{c}$. Quantifying how the transverse width spanned by these concentration gradients across the medium varies with $x$ (\ref{fig:SI-dispersionImages}) then yields a measure of the transverse dispersion coefficient $D^{\nabla}_{\perp}$.

In the case of polymer-free solvent, the flow remains laminar at all injection speeds tested ($\mathrm{Re}\lesssim10^{-2}$). The two streams of fluid initially meet along a single lamellar surface along the channel centerline, characterized by a sharp concentration gradient, that is only minimally stretched and folded by LCA through the disordered pore space (Fig.~\ref{fig:mix}B,D and SI Videos 1--2). Mixing between the streams is therefore minimal. Indeed, we find a large, flow speed-independent value of $l_\mathrm{mix}\approx270d_p$, consistent with previous studies of LCA~\cite{perkins1963review,le2013stretching,le2015lamellar,lester2016chaotic,kree2017scalar,heyman2020stretching}. We also find a small, flow speed-independent value of $D^{\nabla}_{\perp}\approx 0.01Ud_p$, as predicted for spherical grain packings~\cite{perkins1963review,kree2017scalar}. These measurements are shown by the blue upward triangles in Figs.~\ref{fig:mix}F--G, respectively, which we report as a function of $\mathrm{Pe}$, as is typically done.

Adding dilute polymer to the fluid dramatically changes this behavior. In this case, the flow in some pores becomes unstable, with chaotic spatiotemporal fluctuations, at a critical value of $U=U_{c,\mathrm{min}}=170~\muup\mathrm{m/s}$ (\ref{fig:SI-kymographs})---reflecting the onset of EI~\cite{browne2021patches}. Following previous work~\cite{burghelea2007elastic,scholz2014enhanced,kumar2023lagrangian,kumar2023stress,walkama2020disorder,haward2021stagnation,browne2021patches}, we characterize this transition to unstable flow using the Weissenberg number $\mathrm{Wi}=N_1(\dot{\gamma})/\left(2\sigma(\dot{\gamma})\right)$, where the characteristic interstitial shear rate $\dot{\gamma}\equiv U/\sqrt{k/\phi}$ and $k$ is the permeability of the medium; this dimensionless parameter compares the relative strength of flow-induced elastic stresses arising from polymer stretching, quantified by the first normal stress difference $N_1$, to viscous stresses, quantified by the shear stress $\sigma$. The transition to unstable flow then begins in the porous medium at $\mathrm{Wi}=\mathrm{Wi}_{c,\mathrm{min}}=2.6$. As $\mathrm{Wi}$ increases above this value, more and more pores become unstable, until all the pores imaged are unstable at $\mathrm{Wi}=\mathrm{Wi}_{c,\mathrm{max}}=4.4$, as we found previously~\cite{browne2021patches}. 

In this fully-unstable case, the EI-generated flow fluctuations greatly enhance solute mixing, as shown in Figs.~\ref{fig:mix}C,E and SI Videos 3--4. Indeed, as $\mathrm{Wi}$ increases above $\mathrm{Wi}_{c,\mathrm{min}}$, the mixing length $l_\mathrm{mix}$ begins to decrease below its laminar value, eventually saturating at $l_\mathrm{mix}\approx80d_p$ for $\mathrm{Wi}>\mathrm{Wi}_{c,\mathrm{max}}$, as shown by the red downward triangles in Fig.~\ref{fig:mix}F. The transverse dispersion coefficient $D^{\nabla}_{\perp}$ concomitantly increases and saturates at $D^{\nabla}_{\perp}\approx0.06Ud_p$, as shown by the red downward triangles in Fig.~\ref{fig:mix}G. Thus, EI gives rise to a threefold reduction in the length required to appreciably mix solutes in a porous medium, with a corresponding sixfold increase in the transverse dispersivity.\\

\begin{figure*}
    \centering
    \includegraphics[width=\textwidth]{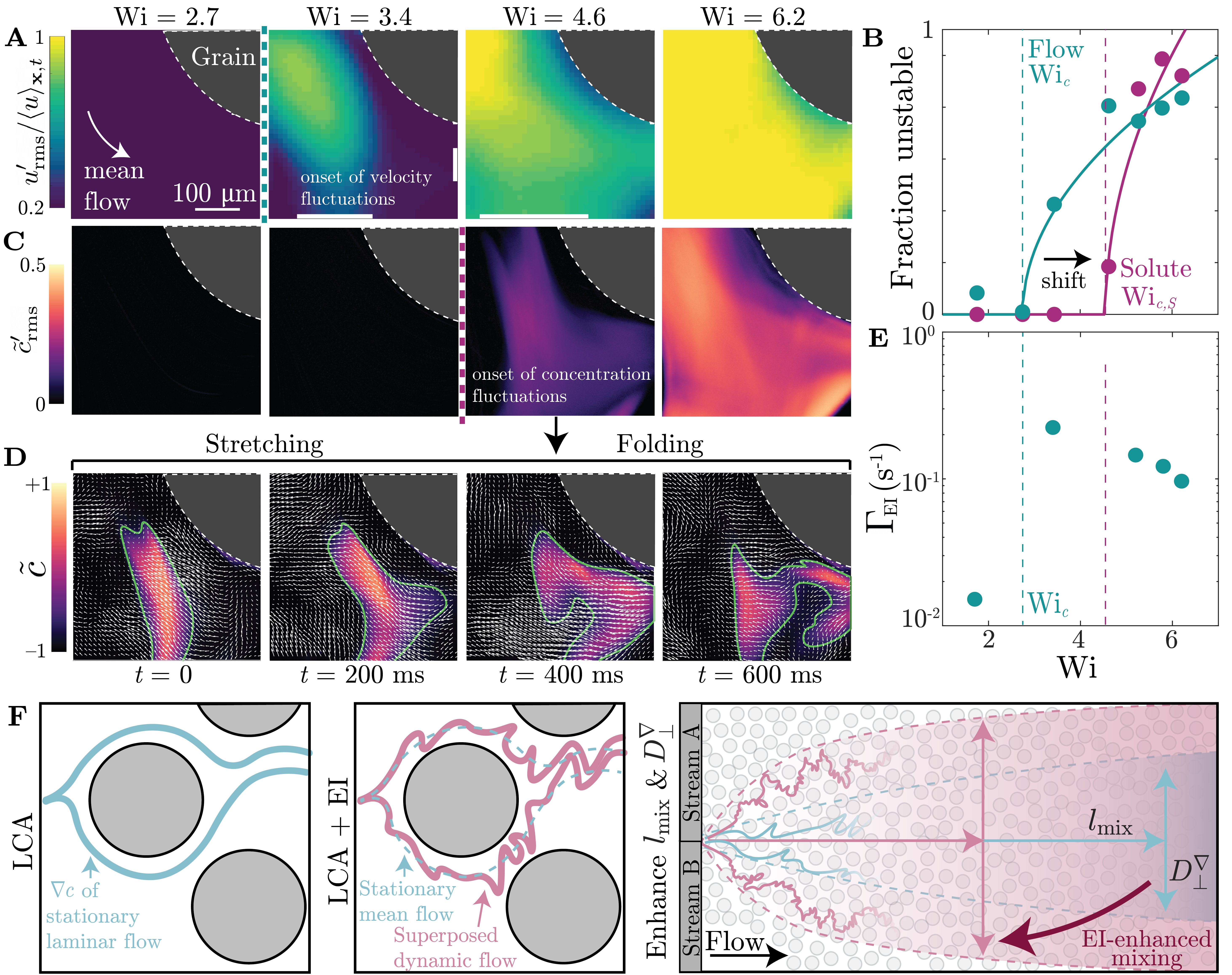}
    \caption{\textbf{EI-generated flow fluctuations stretch and fold solute lamellae in individual pores, enhancing mixing.} 
    \textbf{A,C,D} show magnified views of confocal $x-y$ slices taken at a fixed $z$ position in a single pore during polymer solution flow. We use fluorescent nanoparticles as tracers of the flow, enabling us to map the $x-y$ velocity field $u(\mathbf{x},t)$ via particle image velocimetry. \textbf{A} Map of the root mean square velocity fluctuations $u'_{\mathrm{rms}}(\mathbf{x})$, normalized by the mean $\langle u\rangle_{\mathbf{x},t}$, showing the onset of velocity fluctuations at a critical Weissenberg number $\mathrm{Wi}_c\approx2.9$. The corresponding fluctuations in normalized solvent concentration $c'_{\mathrm{rms}}(\mathbf{x})$ are shown in \textbf{C}; the onset of these fluctuations is shifted to a larger $\mathrm{Wi}_{c,S}\approx4.2$. \textbf{B} The fraction of imaging time that the pore exhibits unstable velocity or solute fluctuations for increases continuously with $\mathrm{Wi}$ above $\mathrm{Wi}_c$ or $\mathrm{Wi}_{c,S}$, as shown by the teal and magenta points, respectively. The curves show power-law fits used to determine $\mathrm{Wi}_c$ and $\mathrm{Wi}_{c,S}$. \textbf{D} Snapshots at four different times for the $\mathrm{Wi}=4.6$ case, showing the normalized solute concentration $\tilde{c}$ along with the fluid velocity field shown by superposed arrows. The EI-generated chaotic flow within the pore stretches and fold the solute blob, increasing its surface area and thereby promoting mixing at the sub-pore scale. \textbf{E} Maximal finite-time Lyapunov exponent of the chaotic flow field in the pore, $\Gamma_\mathrm{EI}$, abruptly increases at $\mathrm{Wi}_c$, reflecting the faster stretching and folding by the fluid after the onset of EI. \textbf{F} Schematic of flow and mixing. (Left) In the case of polymer-free solvent, laminar chaotic advection (LCA) slightly stretches and folds solute lamellae as they transverse multiple disordered pores in the tortuous pore space (blue lines). (Middle) In the unstable case, EI-generated chaotic flow causes additional stretching and folding of solute lamellae within individual pores (red lines). (Right) The combination of LCA and EI results in a reduced mixing length and increased transverse dispersion. }
    \label{fig:Batchelor}
\end{figure*}

\noindent\textbf{Enhanced mixing results from the combination of of laminar chaotic advection across many pores and EI-generated chaotic mixing in individual pores.} Having established that EI can greatly enhance solute mixing in a porous medium, we next ask: What exact features of the unstable flow field generated by EI give rise to this enhanced mixing? And how can this link between fluid flow and solute mixing in a porous medium be described quantitatively? Close inspection of Figs.~\ref{fig:mix}C,E and SI Videos 3--4 provides a clue: EI-generated flow fluctuations appear to successively stretch and fold lamellae of large solute concentration gradient within the pores, increasing the contact area between solute and fluid, thereby promoting mixing via molecular diffusion at small scales. This process is reminiscent of the chaotic mixing generated by inertial turbulence~\cite{shraiman2000scalar,sreenivasan2019turbulent,kumar2023lagrangian}. 

To more quantitatively characterize this phenomenon, we simultaneously image the fluid velocity and solute concentration fields, $\mathbf{u}\left(\mathbf{x},t\right)$ and $\tilde{c}\left(\mathbf{x},t\right)$, respectively, and thereby determine the root mean square fluctuations in the magnitudes of these quantities (indicated by $'_\text{rms}$), in individual pores. A representative example is shown in Fig.~\ref{fig:Batchelor}A--D. As expected for EI, the flow is laminar and stable at small $\mathrm{Wi}$; by contrast, at larger $\mathrm{Wi}>\mathrm{Wi}_c\approx2.9$, we observe intermittent bursts of unstable flow (Fig.~\ref{fig:Batchelor}A, SI Video 5) that increasingly persist with increasing $\mathrm{Wi}$ (teal points in Fig.~\ref{fig:Batchelor}B). These velocity fluctuations do not have any characteristic spatial or temporal scales. Instead, their power spectra decay as power laws, characteristic of chaotic flows (\ref{fig:SI-kymographs})~\cite{browne2021patches}. 

These velocity fluctuations drive fluctuations in solute concentration, which are therefore also chaotic (\ref{fig:SI-kymographs}), with their onset shifted to a larger value of $\mathrm{Wi}=\mathrm{Wi}_{c,S}\approx4.2$ (purple points in Fig.~\ref{fig:Batchelor}B)---reflecting the fact that the solute stream must first enter a pore for it to become mixed by the unstable flow therein. In this case, ``blobs'' of solute (outlined in green in Fig.~\ref{fig:Batchelor}C) are progressively stretched and folded by the chaotic flow in the pore (velocity vectors shown by the arrows in Fig.~\ref{fig:Batchelor}C), exponentially increasing their area over time (\ref{fig:ftle}). Following previous work~\cite{ottino1982description,ottino1989kinematics}, we estimate the characteristic rate of stretching by directly computing the maximal finite-time Lyapunov exponent, $\Gamma_\mathrm{EI}$, from the measured fluid velocity field~\cite{shadden2006lagrangian,peng2009transport}. As shown in Fig.~\ref{fig:Batchelor}E, this characteristic stretching rate increases abruptly at the onset of EI, saturating at $\Gamma_\mathrm{EI}\approx0.2~\text{s}^{-1}\approx0.1\lambda$, where $\lambda$ is the characteristic polymer relaxation time---concomitant with the rapid decrease in $l_\mathrm{mix}$ and increase in $D_\perp^\nabla$ shown in Fig.~\ref{fig:mix}. Thus, EI-generated flow fluctuations promote solute mixing within individual pores in a manner analogous to inertial turbulence, consistent with the findings of previous studies in simpler geometries~\cite{groisman2001,burghelea2004mixing,burghelea2004chaotic,burghelea2007elastic,kumar2023lagrangian,kumar2023stress}.

\begin{figure*}
    \centering
    \includegraphics{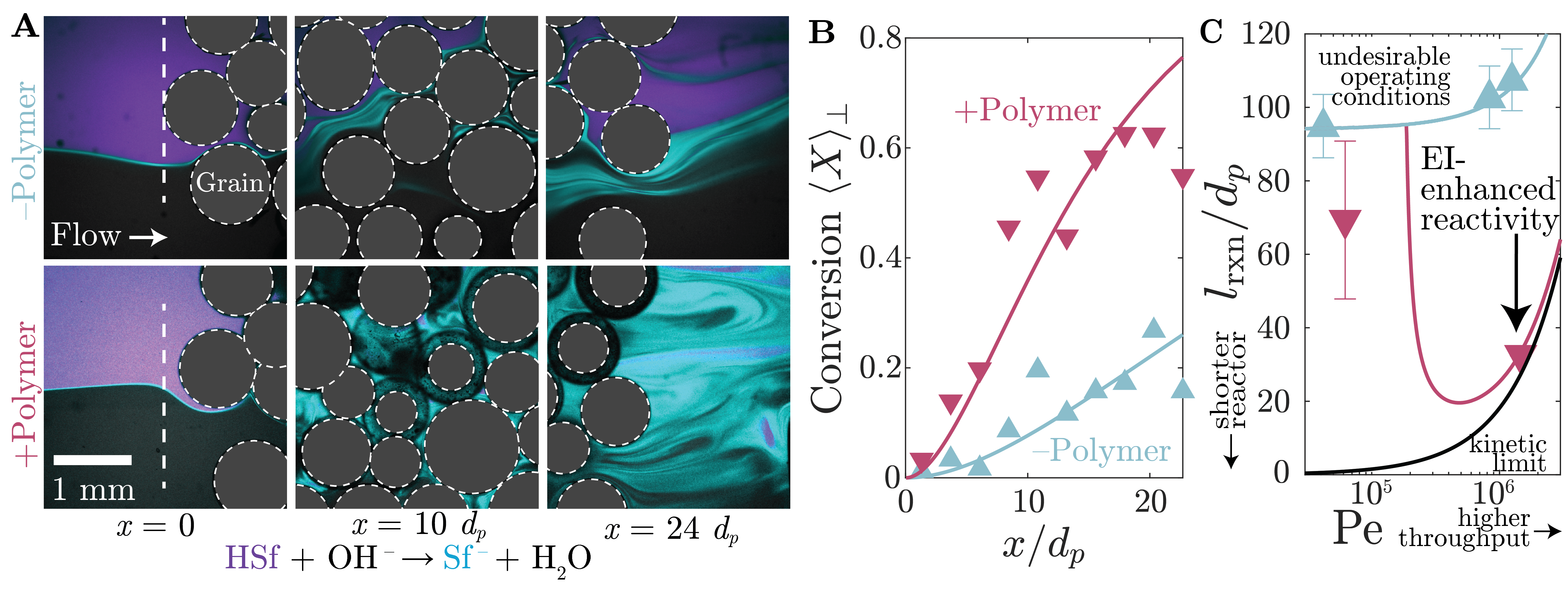}
    \caption{\textbf{EI-generated flow fluctuations enhance chemical reaction kinetics}. \textbf{A} Confocal $x-y$ slices taken at a fixed $z$ position at $x=0$, $10d_p$, and $24d_p$, showing the polymer-free laminar case (top row) or the unstable case of polymer solution (bottom row) at $U=1.3$ mm/s. Stream A contains the phenolic reactive dye $\mathrm{HSf}$ (magenta), while stream B contains the reactant $\mathrm{OH}^-$ along with non-reactive fluorescein dye (grey); the reaction between the two irreversibly generates the product $\mathrm{Sf}^{-}$ (cyan). \textbf{B} The fraction of reactant that has been converted into product $X$, averaged temporally and over the transverse direction $y$, as a function of distance $x$ along the length of the medium. Reaction progress is minimal in the polymer-free laminar case (blue upward triangles), but approaches full conversion more rapidly in the unstable case of polymer solution (red  downward triangles). Curves show Eq.~\ref{eqMain:conversion} with $\tau_k=11\pm1~\mathrm{s}$ and $\theta=0.14\pm0.01$. \textbf{C} Extrapolating the curves in \textbf{B} yields the length of porous medium $l_\mathrm{rxn}$ required to reach a target conversion of $\langle X\rangle_\perp=0.9$; error bars on the points represent uncertainty in determining these values from the fits. In the typical polymer-free laminar case, increasing throughput (increasing $\mathrm{Pe}$) requires a longer reactor (blue upward triangles). EI generated by polymers breaks this trade-off between increasing throughput and decreasing reactor length (red downward triangles). Curves show theoretical predictions obtained using our lamellar mixing model (\emph{Methods}).}
    \label{fig:rxn}
\end{figure*}

Motivated by these observations, we hypothesize that enhanced mixing arises from the combination of two different processes operating at different length and time scales in tandem: laminar chaotic advection (LCA) that stretches and folds solute lamellae at a rate $\Gamma_\mathrm{LCA}\approx U/\left(270d_p\right)$~\cite{perkins1963review,le2013stretching,le2015lamellar,lester2016chaotic,kree2017scalar,heyman2020stretching} as they are transported across multiple disordered pores (Fig.~\ref{fig:Batchelor}F, left panel), and the additional stretching and folding of lamellae within individual pores by the chaotic EI flow field at a rate $\Gamma_\mathrm{EI}\approx0.1\lambda$ (Fig.~\ref{fig:Batchelor}F, middle panel). The rate at which solute becomes appreciably mixed as it traverses the pore space is then given by the superposition of the two:
\begin{equation}\label{eq:main-tauMix}
\tau_\mathrm{mix}^{-1}\approx\Gamma_\mathrm{LCA}+f\Gamma_\mathrm{EI},
\end{equation}
where $\tau_\mathrm{mix}^{-1}\equiv U/l_\mathrm{mix}$, $f$ represents the fraction of pores in which the solute is mixed by EI (\ref{fig:SI-kymographs}), and $\Gamma_\mathrm{EI}=0$ in the purely-laminar polymer-free case. The transverse dispersion coefficient can then be estimated as $D^{\nabla}_{\perp}\approx c d_p^2/\tau_\mathrm{mix}$, with the constant $c\approx3$ given by our measurements in the polymer-free case. Comparing to the measurements shown in Figs.~\ref{fig:mix}F--G provides a direct test of this picture. Remarkably, we find that---despite its simplicity---Eq.~\ref{eq:main-tauMix} provides an excellent description of of how EI both reduces the length required to mix solutes and increases the transverse dispersivity in a porous medium, as shown by the solid red lines in Figs.~\ref{fig:mix}F--G. This agreement between the theoretical model and our experimental measurements confirms the simple conceptual picture of EI-enhanced mixing schematized in Fig.~\ref{fig:Batchelor}F, thereby providing quantitative guidelines to predict and control this phenomenon in practice.\\

\noindent\textbf{EI-enhanced mixing greatly enhances chemical reaction kinetics and yield.} Can this enhanced mixing be harnessed to improve the kinetics and yield of chemical reactions in porous media? To explore this possibility, we introduce a phenolic reactant, $\mathrm{HSf}$, in stream A at a concentration $[\mathrm{HSf}]_0$ (magenta in Fig.~\ref{fig:rxn}A), along with an excess of sodium hydroxide in stream B (black in Fig.~\ref{fig:rxn}A). When these compounds are mixed, the $\mathrm{HSf}$ is irreversibly reduced to its phenolate form $\mathrm{Sf}^{-}$ (cyan in Fig.~\ref{fig:rxn}A); importantly, their peak fluorescence emission occurs at distinct wavelengths~\cite{zurawik2016revisiting}, enabling us to use the imaging to directly quantify reaction progress by tracking the fraction of reactant that has been converted into product, $X\equiv[\mathrm{Sf}^{-}]/[\mathrm{HSf}]_0$ (detailed in \emph{Methods}).

In the polymer-free case, laminar chaotic advection only provides slight mixing between the solvent streams, and the reaction proceeds only along the midplane of the medium, shown by the top row in Fig.~\ref{fig:rxn}A and SI Video 6. As a result, only $\lesssim20\%$ of the reactant is converted to product, as shown by the blue upward triangles in Fig.~\ref{fig:rxn}B. The required porous medium length to reach a target conversion of $90\%$, $l_\mathrm{rxn}$, is then $\gtrsim100d_p$ and \emph{increases} with increasing $\mathrm{Pe}$ (blue upward triangles in Fig.~\ref{fig:rxn}C)---reflecting the typical trade-off between increasing throughput and reducing medium length in porous flow reactors. In stark contrast, the enhanced mixing generated by EI enables the reaction to proceed over a broader region of the pore space (bottom row in Fig.~\ref{fig:rxn}A and SI Video 7), resulting in $\gtrsim60\%$ conversion along the same medium length and a $\sim3\times$ reduction in $l_\mathrm{rxn}$ (red downward triangles in Figs.~\ref{fig:rxn}B--C, respectively). Notably, $l_\mathrm{rxn}$ \emph{decreases} with increasing $\mathrm{Pe}$, indicating that EI-generated mixing helps overcome the trade-off between increasing throughput and reducing reactor length.

Our model of enhanced mixing by successive stretching and folding of solute lamellae both across multiple pores as well as in individual pores, schematized in Fig.~\ref{fig:Batchelor}F, provides a way to rationalize these improvements in reaction kinetics. In particular, considering the successive mixing and reaction of solutes in differential parcels of fluid~\cite{baldyga1999turbulent} yields a prediction for the macroscopic conversion (detailed in \emph{Methods}):
\begin{equation}\label{eqMain:conversion}
    \langle X\rangle_\perp(x)=1-\frac{\mathrm{Da}}{\mathrm{Da}-1}\mathrm{exp}\left(\frac{-x}{U\tau_k\mathrm{Da}}\right)+\frac{1}{\mathrm{Da}-1}\mathrm{exp}\left(\frac{-x}{U\tau_k}\right),
\end{equation}
where the angle brackets indicate an average over the transverse direction $y$ (denoted by the subscript $\perp$), the Damk{\"o}hler number $\mathrm{Da}\equiv\theta\tau_\mathrm{mix}/\tau_k$ compares the time scales of reactant mixing and reaction, $\theta\approx0.14$ is a constant scaling factor to account for reaction initiation before fluid is fully mixed, and $\tau_k=11\pm1$ s, consistent with independent measurements~\cite{guilbert2021chemical}. This prediction is in excellent agreement with the experimental measurements, as shown by the red curve in Fig.~\ref{fig:rxn}B. It also predicts how EI-enhanced mixing enables \emph{both} increased throughput and a reduced reactor length, as shown by the red curve in Fig.~\ref{fig:rxn}C.\\

\noindent\textbf{Discussion.} In many energy, environmental, and industrial processes, solute mixing and chemical reaction kinetics are limited by slow diffusion during laminar flow in a porous medium. Here, we established that elastic instabilities (EI) can be harnessed to overcome this limitation. This new capability simply relies on the addition of dilute, flexible polymers to the fluid, and is therefore straightforward to implement in diverse settings and across a wide range of geometries. It therefore complements other approaches to overcoming the diffusive limitations of typically-laminar flows, which rely on changing the geometry of the boundaries surrounding the fluid instead~\cite{song2003experimental,song2003millisecond,teh2008droplet,i2011droplet,guo2012droplet,shang2017emerging}.

Our in situ imaging revealed that enhanced mixing arises from the successive stretching and folding of solute lamellae by the EI-generated chaotic flow within individual pores, in addition to their stretching and folding as lamellae traverse multiple tortuous pores, in a manner reminiscent of inertial turbulence. As a result, the considerable decrease in the length of medium required to mix solute, increase in the solute transverse dispersivity, and increase in chemical reaction yield imparted by EI can all be predicted using well-established ideas from studies of turbulent flows. We anticipate that the quantitative principles we thereby developed will provide a way to predict and control EI-enhanced mixing and reaction kinetics in applications ranging from large-scale chemical production to environmental remediation~\cite{maree1985biological,maree1987biological,tennakoon1996electrochemical,ajmera2001microfabricated,noorman2007packed,saito1992high,trapp2006unified,saitoh2008column,meirer2018spatial,kudaibergenov2020flow,braff2013membrane,saito1992high,trapp2006unified,saitoh2008column,ajmera2001microfabricated,meirer2018spatial,kudaibergenov2020flow,newman2013role,jensen2017flow,trojanowicz2020flow,vaccaro2017sustainable,trojanowicz2020flow,marre2010synthesis,smith2008compatibility,kitanidis2012delivery,gomez2015denitrification,datta2009redox,matter2016rapid,szulczewski2012lifetime,stegen2016groundwater,bochet2020iron,borer2018spatial,drescher2013biofilm,anglin2008porous,maree1985biological,maree1987biological,tennakoon1996electrochemical,noorman2007packed,smith2008compatibility,anglin2008porous,datta2009redox,kitanidis2012delivery,szulczewski2012lifetime,drescher2013biofilm,gomez2015denitrification,matter2016rapid,stegen2016groundwater,borer2018spatial,bochet2020iron,kumar2022transport}. They also motivate additional studies of the fascinating interplay between viscoelastic fluid dynamics, solute transport and mixing, and chemical reactions.

 Indeed, while here we focused on a simple first-order chemical reaction as a proof of principle, our approach could be adapted to enhance the progression of other types of chemical reactions, such as reactions of other orders, reversible reactions, and solid surface-mediated reactions~\cite{aquino2023fluid}, in porous media. EI-enhanced mixing could also be used to improve heat transfer from a porous medium to an interstitial fluid, such as in heat exchangers and geothermal energy extraction. Finally, while the polymer molecules we used are chemically inert---and thus do not participate in the reaction---it would be interesting to extend our approach to the case where the polymers themselves are reactants or products, potentially driving a more complex feedback between the underlying flow and reaction rates. \\

\noindent\textbf{Acknowledgements.} It is a pleasure to acknowledge stimulating discussions with Holger Stark and Reinier van Buel, and funding support from Princeton University's Materials Research Science and Engineering Center under NSF Grant No. DMR-2011750, as well as the Camille Dreyfus Teacher-Scholar Program. C.A.B. was also supported in part by a Wallace Memorial Honorific Fellowship from the Graduate School at Princeton University.\\

\noindent\textbf{Author contributions. }{C.A.B. and S.S.D. designed the experiments; C.A.B. performed all experiments; C.A.B. and S.S.D. designed the theoretical model; C.A.B. and S.S.D. analyzed all data, discussed the results and implications, and wrote the manuscript; S.S.D. designed and supervised the overall project.}\\

\noindent\textbf{Conflict of interest statement. }{The method described in this publication is the subject of a patent application filed by Princeton University on behalf of C.A.B. and S.S.D.}\\

\section{METHODS} 

\noindent\textbf{Porous media preparation and characterization.} We prepare disordered 3D porous media following our previous work~\cite{krummel2013visualizing,datta2013a,datta2013drainage,browne2021patches,browne2023homogenizing} by densely packing spherical borosilicate glass beads with diameters $d_p=1000$ to $1400~\muup\mathrm{m}$ (\textit{Sigma Aldrich}) in quartz capillaries with rectangular cross sections of area $A=W\times H$, where the width $W=4~\mathrm{mm}$ and height $H=2~\mathrm{mm}$ (\textit{Vitrocom}). Each medium is lightly sintered for $\approx3~\mathrm{min}$ at $1000\degree\mathrm{C}$ to prevent bead rearrangements during flow, and has a length $L=14.7\pm0.1~\mathrm{mm}$ and porosity $\phi\sim0.4$. We affix two inlets and two outlets from bent 14 gauge needles (\textit{McMaster-Carr}), whose outer diameters fit snugly into the rectangular cross-section of the capillary, glued into place with water-tight marine weld (\textit{J-B Weld}). The inlet needles are positioned $\sim1~\mathrm{mm}$ away from the medium to minimize inlet and outlet effects. We determine the medium permeability $k=624\pm3~\muup\mathrm{m^2}$ by injecting the polymer-free solvent at several flow rates $Q=4$ to $40~\mathrm{mL/hr}$, measuring the fully-developed pressure drop $\Delta P$ across the medium using an \textit{Omega PX26} differential pressure transducer and fitting to Darcy's Law $\Delta P/L = \mu Q /\left(kA\right)$.

Before each experiment, we infiltrate the porous medium with isopropyl alcohol (IPA) to prevent trapping of air and then displace the IPA by flushing with water. We then displace the water with the miscible test fluid---either the polymer solution or the polymer-free solvent. We inject our test fluids through both inlets equally at a constant total volumetric flow rate $Q$ ranging from $0.5$ to $25~\mathrm{mL/hr}$ using a \textit{Harvard Apparatus} PHD 2000 syringe pump, maintaining a constant $Q$ for each flow rate tested over $2.5$ hours, corresponding to over $1000$ pore volumes $t_{PV}\equiv \phi AL/Q$, before any micrographs are collected, to ensure an equilibrated state.  At the conclusion of the experiment, we inject rhodamine-dyed polymer-free solvent for 6 hours to fully saturate the pore space with dye, and use confocal microscopy to image the pore space; the binarized images isolate the solid matrix of the porous medium, which we omit from all subsequent image analyses. \\

\noindent\textbf{Fluid preparation and characterization.} All of our test fluids are comprised of a viscous solvent composed of 6 wt.\% ultrapure \textit{milliPore} water, 82.6 wt.\% glycerol (\textit{Sigma Aldrich}), 10.4 wt.\% dimethylsulfoxide (\textit{Sigma Aldrich}), 1 wt.\% NaCl, and $<0.1\%$ additional solutes (specified further below). This formulation precisely matches the fluid refractive index to that of the glass beads $n=1.479$, thus rendering each porous medium transparent when saturated. 

We characterize the fluid rheology using steady shear measurements in an \textit{Anton Paar} MCR301 rheometer, equipped with a 1\degree~5 cm-diameter conical geometry set at a 50 $\muup$m gap over a range of imposed constant shear rates $\dot{\gamma}=0.01$ to $10~\mathrm{s}^{-1}$. The polymer-free solvent is Newtonian, with a constant dynamic viscosity $\mu_s=230~\mathrm{mPa\cdot s}$. The polymer solution has 300 ppm of 18 MDa partially hydrolyzed polyacrylamide (\textit{Polysciences}) added. As detailed previously~\cite{browne2021patches}, this solution is dilute, with an overlap concentration $c^{*}\approx0.77/[\eta]=600\pm300~\mathrm{ppm}$, and the shear stress $\sigma\left(\dot{\gamma}_I\right)=A_s\dot{\gamma}^{\alpha_s}$ and first normal stress difference $N_1\left(\dot{\gamma}_I\right)=A_n\dot{\gamma}^{\alpha_n}$ with $A_s=0.3428\pm0.0002~\mathrm{Pa\cdot s}^{\alpha_s}$, $\alpha_s=0.931\pm0.001$, $A_n=1.16\pm0.03~\mathrm{Pa\cdot s}^{\alpha_n}$, and $\alpha_n=1.25\pm0.02$. Moreover, we describe the solution using a single polymer relaxation time  $\lambda\approx\underset{\dot{\gamma}\rightarrow0}{\text{lim}}~\frac{N_1}{2\left(\eta-\eta_s\right)\dot{\gamma}^2}=480\pm30~\text{ms}$~\cite{shaqfeh1996}. This value is in good agreement with previously reported relaxation times for similar polymer and solvent compositions~\cite{groisman2000,galindo2012,clarke2016,pan2013,clarke2016,walkama2020disorder}, although we expect the true longest relaxation time of the solution to be larger than this value.\\

\noindent\textbf{Macroscopic imaging of solute mixing.} We use Rhodamine Red-X Succinimidyl Ester 5-isomer dye (\textit{Invitrogen}) as a passive solute introduced into stream A at a dilute concentration $c_A=500~\mathrm{ppb}$; stream B is dye-free and hence $c_B=0$. The solute diffusivity is given by $\mathcal{D}\approx1.2\times10^{-8}~\mathrm{cm^2/s}$, obtained by extending previous measurements~\cite{gendron2008diffusion} to the case of our more viscous solvent via the Stokes-Einstein relation. We then image the solute distribution in the pore space using a \textit{Nikon} A1R+ laser scanning confocal fluorescence microscope, continuously acquiring successive $x-y$ optical slices every $2~\mathrm{s}$ at a fixed depth $z\approx600\muup\mathrm{m}$ at 10 different $x$ positions spanning the entire length of the medium. A linear calibration curve obtained using the same imaging settings enables us to then convert fluorescence intensity to dye concentration. 

To determine $l_\mathrm{mix}$, we measure the time-averaged fraction of fluid pixels along $y$ for which $|\tilde{c}|<0.5$ at each position $x$. This fraction of well-mixed fluid increases with position as $1-\mathrm{exp}\left(-x/l_\mathrm{mix}\right)$; fitting this relation to our data then yields the values of $l_\mathrm{mix}$ presented in Fig.~\ref{fig:mix}F.

To determine $D_{\perp}^{\nabla}$, we use the micrographs to determine a map of the solute concentration gradient $|\nabla\tilde{c}|$, as shown in \ref{fig:SI-dispersionImages}. At each position $x$ along the length of the medium, we then quantify the transverse width $\delta$ over which $|\tilde{\nabla}\tilde{c}|>\langle|\tilde{\nabla}\tilde{c}|\rangle_\perp$, where the angle brackets indicate an average over the transverse direction $y$ (denoted by the subscript $\perp$). This plume width increases with position as $\delta(x)\approx \delta(x=0)+D^{\nabla}_{\perp}x/\left(UH\right)$; fitting this relation to our data then yields the values of $D_{\perp}^{\nabla}$ presented in Fig.~\ref{fig:mix}G.\\

\noindent\textbf{Pore-scale imaging of solute mixing and fluid velocity fluctuations.} In addition to using Rhodamine Red-X Succinimidyl Ester 5-isomer dye as a passive solute initially in stream A, we visualize the pore-scale fluid velocity field by seeding both streams A and B with a dilute ($5~\mathrm{ppm}$) suspension of $d_t=200~\mathrm{nm}$-diameter carboxylate-functionalized fluorescent polystyrene FluoSpheres$^{TM}$ (\textit{Invitrogen}), which have a tracer P\'eclet number $(Q/A)d_t/\mathcal{D}_t>10^5\gg1$ and thus act as passive tracers of the fluid flow; here, $\mathcal{D}_t=k_BT/3\pi\eta_0d_t=6\times10^{-3}~\muup\mathrm{m}^2/\mathrm{s}$ is the tracer particle diffusivity obtained via the Stokes-Einstein relation. We monitor the flow in individual pores using a \textit{Nikon} A1R+ laser scanning confocal fluorescence microscope, continuously acquiring successive $x-y$ optical slices of both Rhodamine and tracer particle fluorescence at 15 frames per second at a fixed depth $z\approx600\muup\mathrm{m}$. The tracer particle images are binarized and processed using particle-image velocimetry (PIV)~\cite{thielicke2014pivlab} to obtain the $x-y$ fluid velocity field $\mathbf{u}(\mathbf{x},t)$.

We estimate the characteristic rate of stretching of solute lamellae by directly computing the forward finite-time Lyapunov exponent $\Gamma_\text{ftle}\left(\mathbf{x},t\right)$ from the measured velocity field for the example pore in Fig.~\ref{fig:Batchelor}~\cite{shadden2006lagrangian,peng2009transport}; this quantity measures the exponential rate of separation of virtual tracer particles over the course of one polymer relaxation time $\lambda$. The cumulative distribution of $\Gamma_\text{ftle}$ measured throughout the pore space is shown in \ref{fig:ftle}. We then take the $90^{th}$-percentile value as the characteristic maximal finite-time Lyapunov exponent, $\Gamma_\mathrm{EI}$, which we report in Fig.~\ref{fig:Batchelor}E.\\

\noindent\textbf{In situ characterization of chemical reaction kinetics.} As a model chemical reaction, we choose the irreversible reduction of the fluorescent dye 5(6)-carboxy-SNARF-1 (\textit{Fischer Scientific}) from its phenolic form $\mathrm{HSf}$ introduced in stream A at a concentration $[\mathrm{HSf}]_0=5~\muup\mathrm{M}$ to phenolate form $\mathrm{Sf}^{-}$ in the presence of excess base $\mathrm{NaOH}$ introduced in stream B at a concentration $[\mathrm{OH}^{-}]_0=100~\mathrm{mM}$: $\mathrm{HSf} + \mathrm{OH}^{-} \rightarrow \mathrm{Sf}^{-} + \mathrm{H_2O}$. The reactant and product are fluorescent with distinct emission spectra~\cite{zurawik2016revisiting}, enabling us to map out their spatial distributions using confocal microscopy, converting fluorescence intensity to concentration using a calibration curve obtained using the same imaging settings (\ref{fig:SNARFcal}). To visualize the pore space, stream B is additionally seeded with $1~\muup\mathrm{M}$ of fluorescein, which does not participate in any side reactions.\\

\noindent\textbf{Theoretical model for macroscopic conversion of reactant to product.} For a well-mixed differential parcel of fluid (denoted by the subscript WM), the reaction proceeds according to first-order reaction kinetics~\cite{zurawik2016revisiting,guilbert2021chemical,guilbert2021rectify}: $\frac{\mathrm{d}[\mathrm{Sf}^-]}{\mathrm{d}t}=-k[\mathrm{HSf}][\mathrm{OH}^-]
$, where $[\mathrm{OH}^-]\approx[\mathrm{OH}^-]_0$ since it is in great excess. Thus, the kinetic reaction time scale $\tau_k=\left(k[\mathrm{OH}^-]_0\right)^{-1}$. Stoichiometric mass conservation gives $[\mathrm{HSf}]=[\mathrm{HSf}]_0-[\mathrm{Sf}^-]$, and substituting the fraction of reactant that has been converted into product in this well-mixed differential parcel of fluid, $X_{WM}\equiv[\mathrm{Sf}^{-}]/[\mathrm{HSf}]_0$, yields $X_{WM}=1-\mathrm{exp}\left(-t/\tau_k\right)$. However, this reaction can only begin to proceed in a parcel once it becomes sufficiently mixed at some time $t'$. The fraction of the pore space that is sufficiently-mixed is then given by $f_\mathrm{mix}\approx1-\mathrm{exp}\left(-t/\left(\theta\tau_\mathrm{mix}\right)\right)$, where $0<\theta\leq1$ is a constant scaling factor to account for reaction initiation before fluid is fully mixed. Thus, integrating over all well-mixed parcels of fluid gives the macro-scale conversion~\cite{song2003millisecondreaction,baldyga1999turbulent}: $X\left(t\right)=\int_0^t\frac{\mathrm{d} f_\mathrm{mix}}{\mathrm{d} t}\left(t\right)X_\mathrm{WM}\left(t-t'\right)H\left(t-t'\right)\mathrm{d}t'$, where $H$ is the Heaviside function, which ultimately yields Eq.~\ref{eqMain:conversion}. To compare this prediction with our experimental measurements, we first fit the data shown in Fig.~\ref{fig:rxn}B for the three different experiments with polymer-free solvent at three different values of $\mathrm{Pe}$, as shown by the blue lines in Figs.~\ref{fig:rxn}B--C. These give $\tau_k=11\pm1$ s and $\alpha=0.14\pm0.01$ as best fits. We then apply Eq.~\ref{eqMain:conversion}, using these same parameter values and therefore no fitting parameters, in plotting the theoretical curves for the polymer solution data in Figs.~\ref{fig:rxn}B--C.\\

\clearpage

\setcounter{figure}{0}
\renewcommand{\thefigure}{Supplementary Figure \arabic{figure}}
\renewcommand{\thetable}{Supplementary Table \arabic{table}}
\setcounter{page}{1}

\noindent{\large\textbf{Supporting Video Captions}}

\noindent All videos are available at \url{https://github.com/cabrowne/EnhancedMixing}.\\

\noindent \textbf{Video 1}. Map of solute concentration $\tilde{c}$ imaged at a position $x=10d_p$ along the length of the medium, for the case in which polymer-free solvent is injected at $U=1.7$ mm/s from left to right. The solute (yellow) is initially in the lower stream and the upper stream (black) is initially undyed. Colors match those in Fig.~\ref{fig:mix}B--C. Video is sped up $20\times$. Field of view is $3.2\times3.2~\mathrm{mm}^{2}$. The flow is laminar and steady in time, and so the solute mixes only along the lamellar interface where the streams meet (purple).\\

\noindent \textbf{Video 2}. Map of solute concentration gradient $\tilde{\nabla}\tilde{c}$ corresponding to Video 1. Colors match those in Fig.~\ref{fig:mix}D--E. Video is sped up $20\times$. Field of view is $3.2\times3.2~\mathrm{mm}^{2}$. The flow is laminar and steady in time, and so the solute mixes only along the lamellar interface where the streams meet (bright yellow).\\

\noindent \textbf{Video 3}. Map of solute concentration $\tilde{c}$ imaged at the same position as in Videos 1--2, for the case in which polymer-containing solvent is injected at $U=1.7$ mm/s from left to right. Video is sped up $20\times$. Field of view is $3.2\times3.2~\mathrm{mm}^{2}$.In this case, the chaotic flow continuously generates, stretches, and folds solute lamellae, driving mixing (purple).\\

\noindent \textbf{Video 4}. Map of solute concentration gradient $\tilde{\nabla}\tilde{c}$ corresponding to Video 2.  Video is sped up $20\times$. Field of view is $3.2\times3.2~\mathrm{mm}^{2}$. In this case, the chaotic flow continuously generates, stretches, and folds solute lamellae (bright yellow), driving mixing.\\

\noindent \textbf{Video 5}. Map of velocity fluctuations $u'$ in the representative pore shown in Fig~\ref{fig:Batchelor}A. These intermittent bursts of instability arise from EI generated by the polymer solution at $Wi=3.4$. Colors match those in Fig~\ref{fig:Batchelor}A. Video is sped up $6.7\times$. \\

\noindent \textbf{Video 6}. Map of reaction progress imaged at a position $x=10d_p$ along the length of the medium, for the case in which polymer-free solvent is injected at $U=1.7$ mm/s from left to right. The reagents (magenta and grey) form product (cyan) when they mix and react. Colors match those in Fig.~\ref{fig:mix}B--C. Video is sped up $10\times$. Field of view is $3.2\times3.2~\mathrm{mm}^{2}$. The flow is laminar and steady in time, and so the reaction proceeds only along the lamellar interface where the streams meet (cyan).\\

\noindent \textbf{Video 7}. Map of reaction progress imaged at a similar position as in Video 6, for the case in which polymer-containing solvent is injected at $U=1.7$ mm/s from left to right. Video is sped up $10\times$. Field of view is $3.2\times3.2~\mathrm{mm}^{2}$.In this case, the chaotic flow continuously generates, stretches, and folds solute lamellae, driving mixing and chemical reaction progress (cyan).\\

\begin{figure*}
    \centering
    \includegraphics{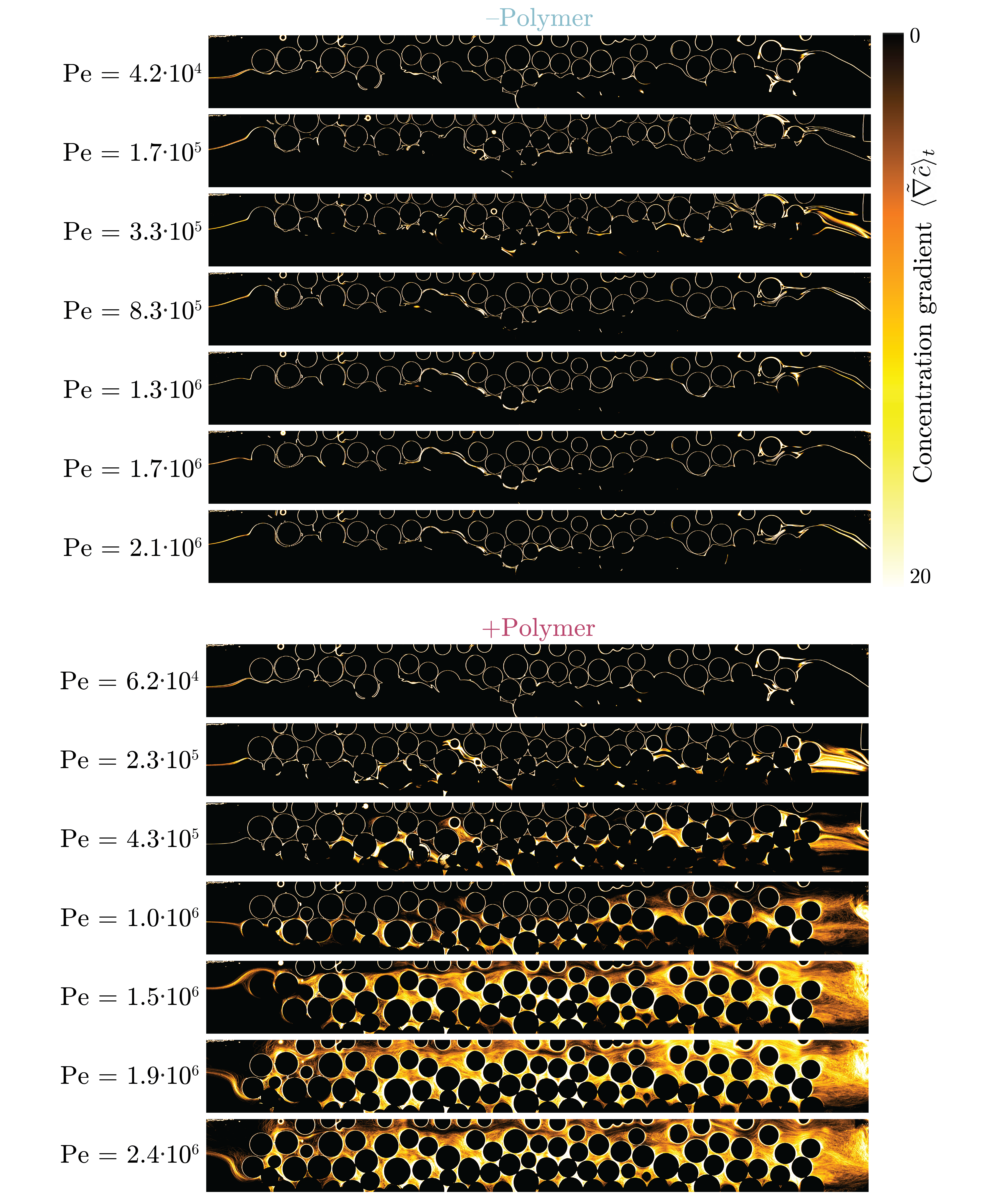}
    \caption{Map of concentration gradients $|\tilde{\nabla}\tilde{c}|$ inside the pore space. The gradient field is time-averaged within each field of view over the 3 min imaging window and stitched to show the fraction of the medium accessed at any point in time. The width of the plume encompassing strong solute gradients increases with the length along the medium, reflecting transverse dispersion. Steep gradients appearing at the surfaces of the solid beads in the top half of the images are an artifact of numerical differentiation.}
    \label{fig:SI-dispersionImages}
\end{figure*}

\begin{figure*}
    \centering
    \includegraphics[height=16cm]{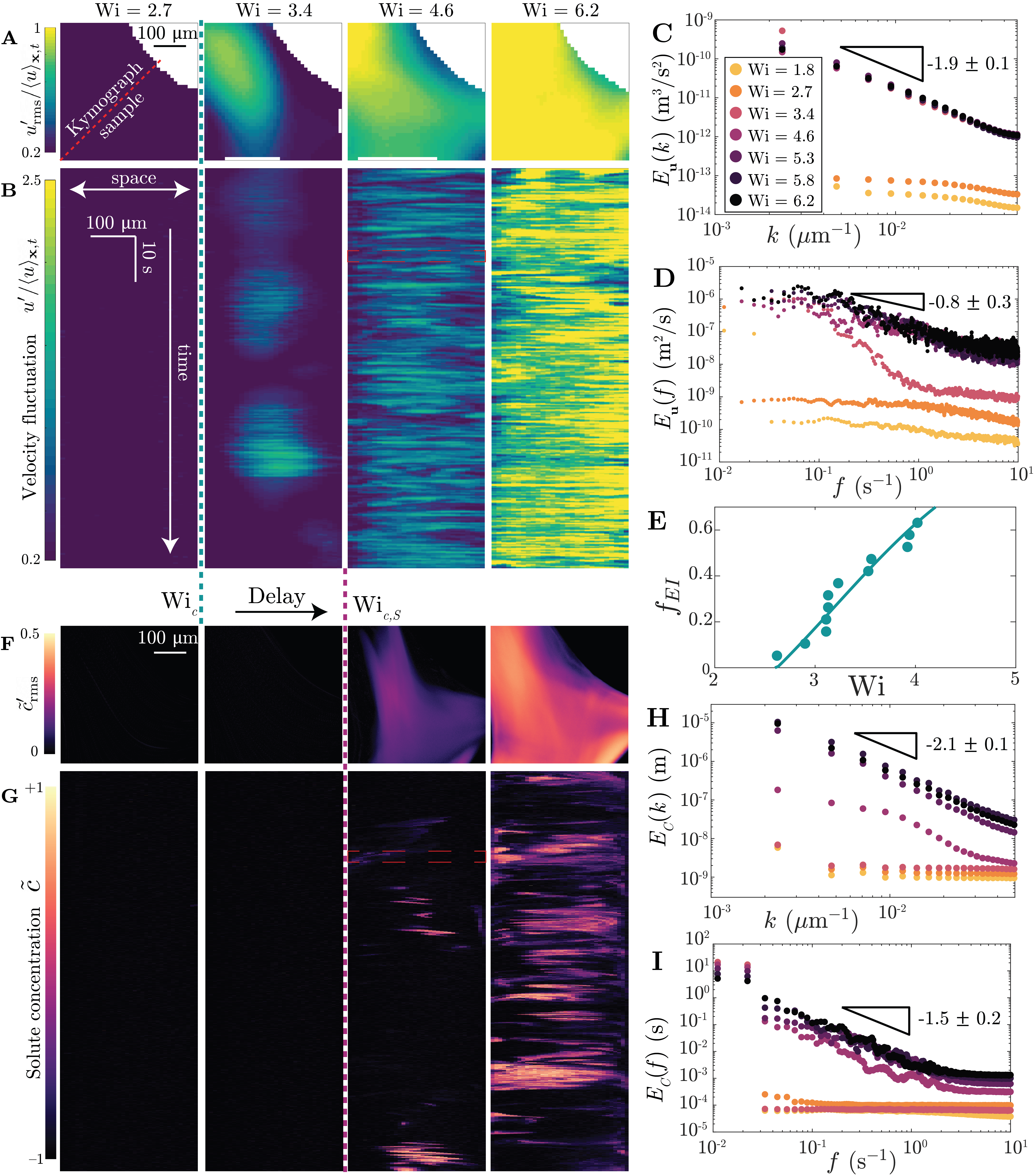}
    \caption{As we found previously~\cite{browne2021patches}, in the case of polymer solution flow, individual pores transition from stable, laminar flow at low $\mathrm{Wi}$ to EI-generated unstable flow at a critical value of $\mathrm{Wi}=\mathrm{Wi}_c$, which we determine from power-law fits as shown in Fig.~\ref{fig:Batchelor}. \textbf{A} We characterize this transition to EI by monitoring the root mean square fluctuations in the velocity magnitude, $u'_{\mathrm{rms}}(\mathbf{x})$, normalized by the spatially- and temporally-averaged velocity magnitude $\langle u\rangle_{\mathbf{x},t}$, in an example pore. \textbf{B} Kymographs corresponding to the pore centerline diagonal indicated in A, showing that intermittent bursts of instability arise and increasingly persist with increasing $\mathrm{Wi}$ above $\mathrm{Wi}_c$. \textbf{C--D} Power spectral density of flow velocity fluctuations of different wavenumbers $k$ and frequencies $f$, computed as the square of the Fourier transform of velocity fluctuations. We observe an abrupt increase when $\mathrm{Wi}$ exceeds $\mathrm{Wi}_c$; the power spectra show power law decays $E_u\left(k\right)\sim k^{-\alpha_u}$ and $E_u\left(f\right)\sim f^{-\beta_u}$ indicating that the flow is chaotic, with $\alpha_u\approx1.9$ and $\beta_u\approx0.8$. While different forms of EI may arise in different geometries, possibly resulting in flows with distinct power spectra of flow velocity fluctuations, the values we measure are consistent with the ranges $1\lesssim\alpha_u\lesssim3$ and $1\lesssim\beta_u\lesssim4$ reported for EI in various other
geometries~\cite{groisman2000,groisman2004elastic,jha2020universal,haward2021stagnation,qin2017characterizing,vanBuel2018elastic,vanBuel2020active,browne2021patches}. \textbf{E} Different pores have different values of $\mathrm{Wi}_c$, presumably reflecting pore-to-pore differences in geometry, as we found previously~\cite{browne2021patches}. This figure shows the fraction of unstable pores $f_{\mathrm{EI}}$ at a given $\mathrm{Wi}$ measured across multiple pores in the medium.  The data are well described by the function $f_\mathrm{EI}\left(\mathrm{Wi}\right)=1-e^{-\omega_\mathrm{EI}\left(\mathrm{Wi}^{\alpha_s/\alpha_n}-\mathrm{Wi}_{c,\mathrm{min}}^{\alpha_s/\alpha_n}\right)}$, where $\mathrm{Wi}_{c,\mathrm{min}}=2.6$ and $\omega_\mathrm{EI}=54$, and $\alpha_s=0.931$ and $\alpha_n=1.25$ are the parameters obtained from shear rheology that relate shear rate $\dot{\gamma}$ to shear stress $\sigma$ and first normal stress difference $N_1$, respectively, thereby providing a mapping between $\dot{\gamma}$ and $\mathrm{Wi}\equiv N1/(2\sigma)$. Injected solute does not disperse throughout the entire pore space, and thus, the fraction of pores in which solute is mixed by EI is given by $f$, where $f\left(\mathrm{Wi}\right)$ is given by same function as $f_\mathrm{EI}\left(\mathrm{Wi}\right)$, with the same values of $\mathrm{Wi}_{c,\mathrm{min}}$, $\alpha_s$, and $\alpha_n$, but with $\omega_{\mathrm{EI}}$ rescaled to $\omega$: $f\left(\mathrm{Wi}\right)=1-e^{-\omega\left(\mathrm{Wi}^{\alpha_s/\alpha_n}-\mathrm{Wi}_{c,\mathrm{min}}^{\alpha_s/\alpha_n}\right)}$, with $\omega=490$ obtained by fitting Eq.~\ref{eq:main-tauMix} to the data in Fig.~\ref{fig:mix}F. \textbf{F--G} We observe corresponding intermittent fluctuations in solvent concentration, shifted to higher values of $\mathrm{Wi}$, reflecting the fact that
the solute stream must first enter a pore for it to become mixed by the unstable flow therein. \textbf{H--I} Power spectra of solute concentration fluctuations also decay as power laws for $\mathrm{Wi}>\mathrm{Wi}_{c,S}$: $E_c\left(k\right)\sim k^{-\alpha_c}$ and $E_c\left(f\right)\sim f^{-\beta_c}$ , with $\alpha_c\approx2.1$ and $\beta_c\approx1.5$.}
    \label{fig:SI-kymographs}
\end{figure*}

\begin{figure*}
    \centering
    \includegraphics{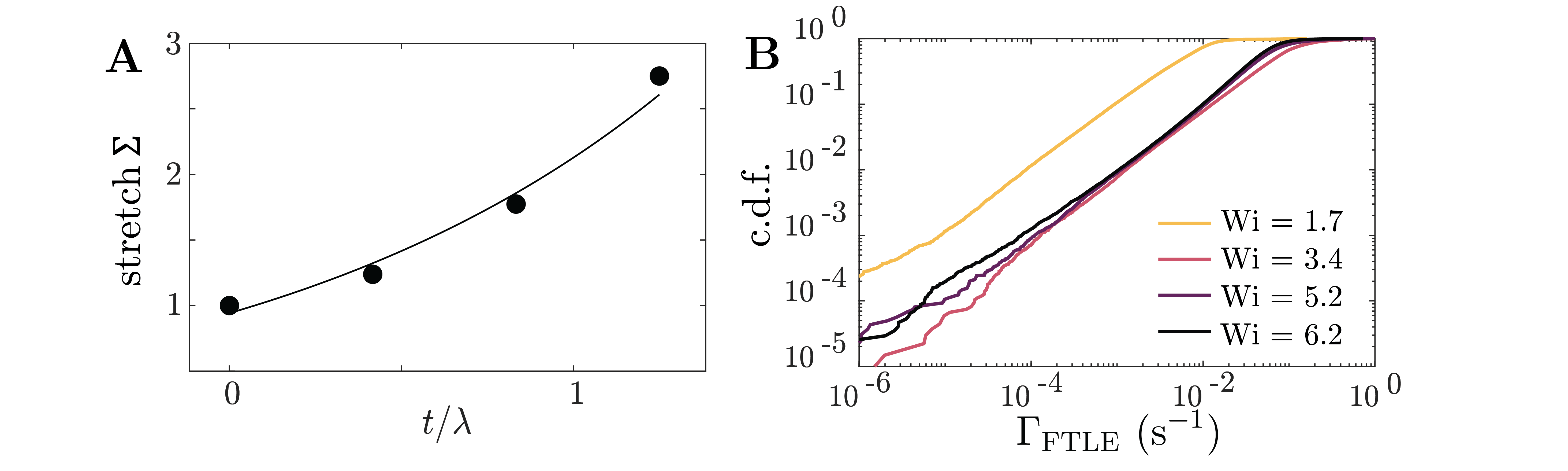}
    \caption{\textbf{A} Surface area of the representative solute blob shown in Fig.~\ref{fig:Batchelor}C increases exponentially in time; here, stretch $\Sigma\equiv \left(\frac{l(t)}{l(t=0)}\right)^2$, where $l(t)$ is the measured blob perimeter. \textbf{B} Cumulative distribution function of the forward finite-time Lyapunov exponent $\Gamma_\text{ftle}\left(\mathbf{x},t\right)$ determined directly from the measured velocity field for the example pore in Fig.~\ref{fig:Batchelor}. }
    \label{fig:ftle}
\end{figure*}

\begin{figure*}
    \centering
    \includegraphics{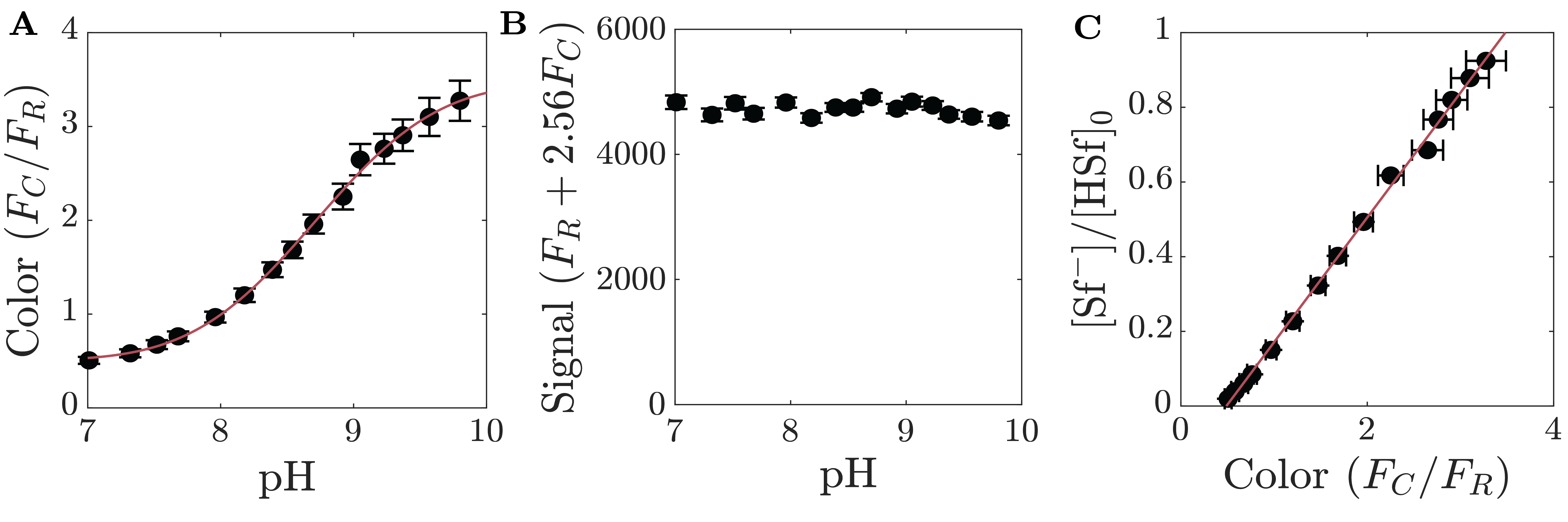}
    \caption{We obtain a calibration curve by which local fluorescence intensity $F_C$ of the phenolate product $\mathrm{Sf}^{-}$ relative to the intensity $F_R$ of the phenolic reactant $\mathrm{HSf}$ can be converted to a measure of local conversion represented by the fraction of reactant that has been converted into product, $[\mathrm{Sf}^{-}]/[\mathrm{HSf}]_0$. To do so, we image calibration standards in empty quartz capillaries using the same imaging settings as in the main text experiments. The standards have $[\mathrm{HSf}]_0=5~\muup\mathrm{M}$ at various intermediate pH values ranging from $7$ to $10$ using $1$ mM Tris-base buffered against hydrochloric acid. At each pH, the reaction reaches an equilibrium defined by its $\mathrm{pK_b}\equiv-\mathrm{log}\left(\mathrm{K_b}\right)$, where $\mathrm{K_b}\equiv\frac{[\mathrm{HSf}][\mathrm{OH}^{-}]}{[\mathrm{Sf}^{-}]}$. Stoichiometric conservation additionally imposes $[\mathrm{HSf}]_0=[\mathrm{HSf}]+[\mathrm{Sf}^{-}]$, and $[\mathrm{OH}]=10^{-\mathrm{pOH}}$ since the solution is buffered. \textbf{A} Measurements of $F_C/F_R$ as a function of pH, which shifts the equilibrium ratio $[\mathrm{Sf}^{-}]/[\mathrm{HSf}]$. We fit an error function to the data (red line), which yields an estimate for the $\mathrm{pK_a}=14-\mathrm{pK_b}=8.7$. \textbf{B} Characterization of the reduction in overall fluorescent intensity at intermediate phenolic/phenolate ratios allows the absolute concentrations to be measured using a best fit line.
    \textbf{C} Together these curves allow the local conversion $X=[\mathrm{Sf}^{-}]/[\mathrm{HSf}]_0$ to be determined directly from local relative fluorescence intensity $F_C/F_R$. The red line indicates a best-fit linear calibration curve. }
    \label{fig:SNARFcal}
\end{figure*}


\begin{thebibliography}{10}

\bibitem{hoover1950re}
Georgius Agricola, Herbert Hoover, and Lou~Henry Hoover.
\newblock {\em De re metallica}.
\newblock Courier Corporation, 1556.

\bibitem{ottino1989kinematics}
Julio~M Ottino.
\newblock {\em The kinematics of mixing: stretching, chaos, and transport},
  volume~3.
\newblock Cambridge university press, 1989.

\bibitem{kree2017scalar}
Mihkel Kree and Emmanuel Villermaux.
\newblock Scalar mixtures in porous media.
\newblock {\em Physical Review Fluids}, 2(10):104502, 2017.

\bibitem{heyman2020stretching}
Joris Heyman, Daniel~R Lester, R{\'e}gis Turuban, Yves M{\'e}heust, and Tanguy
  Le~Borgne.
\newblock Stretching and folding sustain microscale chemical gradients in
  porous media.
\newblock {\em Proceedings of the National Academy of Sciences},
  117(24):13359--13365, 2020.

\bibitem{groisman2001}
Alexander Groisman and Victor Steinberg.
\newblock Efficient mixing at low {R}eynolds numbers using polymer additives.
\newblock {\em Nature}, 410(6831):905, 2001.

\bibitem{burghelea2004chaotic}
Teodor Burghelea, Enrico Segre, Israel Bar-Joseph, Alex Groisman, and Victor
  Steinberg.
\newblock Chaotic flow and efficient mixing in a microchannel with a polymer
  solution.
\newblock {\em Phys. Rev. E}, 69(6):066305, 2004.

\bibitem{burghelea2004mixing}
T~Burghelea, E~Segre, and V~Steinberg.
\newblock Mixing by polymers: Experimental test of decay regime of mixing.
\newblock {\em Phys. Rev. Lett.}, 92(16):164501, 2004.

\bibitem{burghelea2007elastic}
Teodor Burghelea, Enrico Segre, and Victor Steinberg.
\newblock Elastic turbulence in von karman swirling flow between two disks.
\newblock {\em Phys. Fluids}, 19(5):053104, 2007.

\bibitem{maree1985biological}
JP~Maree and Wilma~F Strydom.
\newblock Biological sulphate removal in an upflow packed bed reactor.
\newblock {\em Water Research}, 19(9):1101--1106, 1985.

\bibitem{maree1987biological}
JP~Maree and Wilma~F Strydom.
\newblock Biological sulphate removal from industrial effluent in an upflow
  packed bed reactor.
\newblock {\em Water Research}, 21(2):141--146, 1987.

\bibitem{tennakoon1996electrochemical}
CLK Tennakoon, RC~Bhardwaji, and JO'M Bockris.
\newblock Electrochemical treatment of human wastes in a packed bed reactor.
\newblock {\em Journal of applied electrochemistry}, 26(1):18--29, 1996.

\bibitem{ajmera2001microfabricated}
Sameer~K Ajmera, Matthew~W Losey, Klavs~F Jensen, and Martin~A Schmidt.
\newblock Microfabricated packed-bed reactor for phosgene synthesis.
\newblock {\em AIChE Journal}, 47(7):1639--1647, 2001.

\bibitem{noorman2007packed}
Sander Noorman, Martin van Sint~Annaland, and Hans Kuipers.
\newblock Packed bed reactor technology for chemical-looping combustion.
\newblock {\em Industrial \& Engineering Chemistry Research},
  46(12):4212--4220, 2007.

\bibitem{saito1992high}
Koichi Saito, Mazakazu Horie, Norihide Nose, Kazuya Nakagomi, and Hiroyuki
  Nakazawa.
\newblock High-performance liquid chromatography of histamine and
  1-methylhistamine with on-column fluroescence derivatization.
\newblock {\em Journal of Chromatography A}, 595(1-2):163--168, 1992.

\bibitem{trapp2006unified}
O~Trapp.
\newblock Unified equation for access to rate constants of first-order
  reactions in dynamic and on-column reaction chromatography.
\newblock {\em Analytical chemistry}, 78(1):189--198, 2006.

\bibitem{saitoh2008column}
Kazunori Saitoh, Kohta Koichi, Fumiko Yabiku, Yumiko Noda, Marc~D Porter, and
  Masami Shibukawa.
\newblock On-column electrochemical redox derivatization for enhancement of
  separation selectivity of liquid chromatography: Use of redox reaction as
  secondary chemical equilibrium.
\newblock {\em Journal of Chromatography A}, 1180(1-2):66--72, 2008.

\bibitem{meirer2018spatial}
Florian Meirer and Bert~M Weckhuysen.
\newblock Spatial and temporal exploration of heterogeneous catalysts with
  synchrotron radiation.
\newblock {\em Nature Reviews Materials}, 3(9):324--340, 2018.

\bibitem{kudaibergenov2020flow}
Sarkyt~E Kudaibergenov and Gulzhian~I Dzhardimalieva.
\newblock Flow-through catalytic reactors based on metal nanoparticles
  immobilized within porous polymeric gels and surfaces/hollows of polymeric
  membranes.
\newblock {\em Polymers}, 12(3):572, 2020.

\bibitem{braff2013membrane}
William~A Braff, Martin~Z Bazant, and Cullen~R Buie.
\newblock Membrane-less hydrogen bromine flow battery.
\newblock {\em Nature communications}, 4(1):1--6, 2013.

\bibitem{newman2013role}
Stephen~G Newman and Klavs~F Jensen.
\newblock The role of flow in green chemistry and engineering.
\newblock {\em Green chemistry}, 15(6):1456--1472, 2013.

\bibitem{jensen2017flow}
Klavs~F Jensen.
\newblock Flow chemistry—microreaction technology comes of age.
\newblock {\em AIChE Journal}, 63(3):858--869, 2017.

\bibitem{trojanowicz2020flow}
Marek Trojanowicz.
\newblock Flow chemistry in contemporary chemical sciences: A real variety of
  its applications.
\newblock {\em Molecules}, 25(6):1434, 2020.

\bibitem{vaccaro2017sustainable}
Luigi Vaccaro.
\newblock {\em Sustainable flow chemistry: methods and applications}.
\newblock John Wiley \& Sons, 2017.

\bibitem{marre2010synthesis}
Samuel Marre and Klavs~F Jensen.
\newblock Synthesis of micro and nanostructures in microfluidic systems.
\newblock {\em Chemical Society Reviews}, 39(3):1183--1202, 2010.

\bibitem{lee2016passive}
Chia-Yen Lee, Wen-Teng Wang, Chan-Chiung Liu, and Lung-Ming Fu.
\newblock Passive mixers in microfluidic systems: A review.
\newblock {\em Chemical Engineering Journal}, 288:146--160, 2016.

\bibitem{song2003experimentalmixing}
Helen Song, Michelle~R Bringer, Joshua~D Tice, Cory~J Gerdts, and Rustem~F
  Ismagilov.
\newblock Experimental test of scaling of mixing by chaotic advection in
  droplets moving through microfluidic channels.
\newblock {\em Applied physics letters}, 83(22):4664--4666, 2003.

\bibitem{perkins1963review}
TK~Perkins and OC~Johnston.
\newblock A review of diffusion and dispersion in porous media.
\newblock {\em Society of Petroleum Engineers Journal}, 3(01):70--84, 1963.

\bibitem{le2013stretching}
Tanguy Le~Borgne, Marco Dentz, and Emmanuel Villermaux.
\newblock Stretching, coalescence, and mixing in porous media.
\newblock {\em Physical review letters}, 110(20):204501, 2013.

\bibitem{le2015lamellar}
Tanguy Le~Borgne, Marco Dentz, and Emmanuel Villermaux.
\newblock The lamellar description of mixing in porous media.
\newblock {\em Journal of Fluid Mechanics}, 770:458--498, 2015.

\bibitem{lester2016chaotic}
Daniel~R Lester, Marco Dentz, and Tanguy Le~Borgne.
\newblock Chaotic mixing in three-dimensional porous media.
\newblock {\em Journal of Fluid Mechanics}, 803:144--174, 2016.

\bibitem{guilbert2021chemical}
Emilie Guilbert, Christophe Almarcha, and Emmanuel Villermaux.
\newblock Chemical reaction for mixing studies.
\newblock {\em Physical Review Fluids}, 6(11):114501, 2021.

\bibitem{guilbert2021rectify}
Emilie Guilbert and Emmanuel Villermaux.
\newblock Chemical reactions rectify mixtures composition.
\newblock {\em Physical Review Fluids}, 6(11):L112501, 2021.

\bibitem{smith2008compatibility}
Megan~M Smith, Jeff~AK Silva, Junko Munakata-Marr, and John~E McCray.
\newblock Compatibility of polymers and chemical oxidants for enhanced
  groundwater remediation.
\newblock {\em Environmental science \& technology}, 42(24):9296--9301, 2008.

\bibitem{kitanidis2012delivery}
Peter~K Kitanidis and Perry~L McCarty.
\newblock {\em Delivery and Mixing in the Subsurface: Processes and Design
  Principles for in Situ remediation}, volume~4.
\newblock Springer Science \& Business Media, 2012.

\bibitem{gomez2015denitrification}
Jesus~D Gomez-Velez, Judson~W Harvey, M~Bayani Cardenas, and Brian Kiel.
\newblock Denitrification in the mississippi river network controlled by flow
  through river bedforms.
\newblock {\em Nature Geoscience}, 8(12):941--945, 2015.

\bibitem{datta2009redox}
S~Datta, B~Mailloux, H-B Jung, MA~Hoque, M~Stute, KM~Ahmed, and Y~Zheng.
\newblock Redox trapping of arsenic during groundwater discharge in sediments
  from the meghna riverbank in bangladesh.
\newblock {\em Proceedings of the National Academy of Sciences},
  106(40):16930--16935, 2009.

\bibitem{matter2016rapid}
Juerg~M Matter, Martin Stute, Sandra~{\'O} Sn{\ae}bj{\"o}rnsdottir, Eric~H
  Oelkers, Sigurdur~R Gislason, Edda~S Aradottir, Bergur Sigfusson, Ingvi
  Gunnarsson, Holmfridur Sigurdardottir, Einar Gunnlaugsson, et~al.
\newblock Rapid carbon mineralization for permanent disposal of anthropogenic
  carbon dioxide emissions.
\newblock {\em Science}, 352(6291):1312--1314, 2016.

\bibitem{szulczewski2012lifetime}
Michael~L Szulczewski, Christopher~W MacMinn, Howard~J Herzog, and Ruben
  Juanes.
\newblock Lifetime of carbon capture and storage as a climate-change mitigation
  technology.
\newblock {\em Proceedings of the National Academy of Sciences},
  109(14):5185--5189, 2012.

\bibitem{stegen2016groundwater}
James~C Stegen, James~K Fredrickson, Michael~J Wilkins, Allan~E Konopka,
  William~C Nelson, Evan~V Arntzen, William~B Chrisler, Rosalie~K Chu, Robert~E
  Danczak, Sarah~J Fansler, et~al.
\newblock Groundwater--surface water mixing shifts ecological assembly
  processes and stimulates organic carbon turnover.
\newblock {\em Nature communications}, 7(1):1--12, 2016.

\bibitem{bochet2020iron}
Olivier Bochet, Lorine Bethencourt, Alexis Dufresne, Julien Farasin, Mathieu
  P{\'e}drot, Thierry Labasque, Eliot Chatton, Nicolas Lavenant, Christophe
  Petton, Benjamin~W Abbott, et~al.
\newblock Iron-oxidizer hotspots formed by intermittent oxic--anoxic fluid
  mixing in fractured rocks.
\newblock {\em Nature Geoscience}, 13(2):149--155, 2020.

\bibitem{borer2018spatial}
Benedict Borer, Robin Tecon, and Dani Or.
\newblock Spatial organization of bacterial populations in response to oxygen
  and carbon counter-gradients in pore networks.
\newblock {\em Nature communications}, 9(1):1--11, 2018.

\bibitem{drescher2013biofilm}
Knut Drescher, Yi~Shen, Bonnie~L Bassler, and Howard~A Stone.
\newblock Biofilm streamers cause catastrophic disruption of flow with
  consequences for environmental and medical systems.
\newblock {\em Proceedings of the National Academy of Sciences},
  110(11):4345--4350, 2013.

\bibitem{anglin2008porous}
Emily~J Anglin, Lingyun Cheng, William~R Freeman, and Michael~J Sailor.
\newblock Porous silicon in drug delivery devices and materials.
\newblock {\em Advanced drug delivery reviews}, 60(11):1266--1277, 2008.

\bibitem{vinogradov1965experimental}
GV~Vinogradov and VN~Manin.
\newblock An experimental study of elastic turbulence.
\newblock {\em Kolloid-Zeitschrift und Zeitschrift f{\"u}r Polymere},
  201(2):93--98, 1965.

\bibitem{larson1990}
Ronald~G Larson, Eric~SG Shaqfeh, and Susan~J Muller.
\newblock A purely elastic instability in {T}aylor--{C}ouette flow.
\newblock {\em Journal of Fluid Mechanics}, 218:573--600, 1990.

\bibitem{groisman2000}
Alexander Groisman and Victor Steinberg.
\newblock Elastic turbulence in a polymer solution flow.
\newblock {\em Nature}, 405(6782):53, 2000.

\bibitem{scholz2014enhanced}
Christian Scholz, Frank Wirner, Juan~Ruben Gomez-Solano, and Clemens Bechinger.
\newblock Enhanced dispersion by elastic turbulence in porous media.
\newblock {\em EPL (Europhysics Letters)}, 107(5):54003, 2014.

\bibitem{kumar2023lagrangian}
Manish Kumar, Jeffrey~S Guasto, and Arezoo~M Ardekani.
\newblock Lagrangian stretching reveals stress topology in viscoelastic flows.
\newblock {\em Proceedings of the National Academy of Sciences},
  120(5):e2211347120, 2023.

\bibitem{kumar2023stress}
Manish Kumar, Derek~M Walkama, Arezoo~M Ardekani, and Jeffrey~S Guasto.
\newblock Stress and stretching regulate dispersion in viscoelastic porous
  media flows.
\newblock {\em Soft Matter}, 2023.

\bibitem{damseh2010visco}
Rebhi~A Damseh and BA~Shannak.
\newblock Visco-elastic fluid flow past an infinite vertical porous plate in
  the presence of first-order chemical reaction.
\newblock {\em Applied Mathematics and Mechanics}, 31(8):955--962, 2010.

\bibitem{sasmal2023applications}
C~Sasmal.
\newblock Applications of elastic instability and elastic turbulence: Review,
  limitations, and future directions.
\newblock {\em arXiv preprint arXiv:2301.02395}, 2023.

\bibitem{kumar2022transport}
Manish Kumar, Jeffrey~S Guasto, and Arezoo~M Ardekani.
\newblock Transport of complex and active fluids in porous media.
\newblock {\em Journal of Rheology}, 66(2):375--397, 2022.

\bibitem{walkama2020disorder}
Derek~M Walkama, Nicolas Waisbord, and Jeffrey~S Guasto.
\newblock Disorder suppresses chaos in viscoelastic flows.
\newblock {\em {Physical Review Letters}}, 124(16):164501, 2020.

\bibitem{haward2021stagnation}
Simon~J Haward, Cameron~C Hopkins, and Amy~Q Shen.
\newblock Stagnation points control chaotic fluctuations in viscoelastic porous
  media flow.
\newblock {\em Proceedings of the National Academy of Sciences},
  118(38):e2111651118, 2021.

\bibitem{browne2021patches}
Christopher~A. Browne and Sujit~S. Datta.
\newblock Elastic turbulence generates anomalous flow resistance in porous
  media.
\newblock {\em Science Advances}, 7:eabj2619, 2021.

\bibitem{datta2022perspectives}
Sujit~S Datta, Arezoo~M Ardekani, Paulo~E Arratia, Antony~N Beris, Irmgard
  Bischofberger, Gareth~H McKinley, Jens~G Eggers, J~Esteban L{\'o}pez-Aguilar,
  Suzanne~M Fielding, Anna Frishman, et~al.
\newblock Perspectives on viscoelastic flow instabilities and elastic
  turbulence.
\newblock {\em Physical Review Fluids}, 7(8):080701, 2022.

\bibitem{browne2023homogenizing}
Christopher~A Browne, Richard~B Huang, Callie~W Zheng, and Sujit~S Datta.
\newblock Homogenizing fluid transport in stratified porous media using an
  elastic flow instability.
\newblock {\em Journal of Fluid Mechanics}, 963:A30, 2023.

\bibitem{chertkov2003decay}
Michael Chertkov and Vladimir Lebedev.
\newblock Decay of scalar turbulence revisited.
\newblock {\em Physical review letters}, 90(3):034501, 2003.

\bibitem{son1999turbulent}
DT~Son.
\newblock Turbulent decay of a passive scalar in the batchelor limit: Exact
  results from a quantum-mechanical approach.
\newblock {\em Physical Review E}, 59(4):R3811, 1999.

\bibitem{balkovsky1999universal}
E~Balkovsky and A~Fouxon.
\newblock Universal long-time properties of lagrangian statistics in the
  batchelor regime and their application to the passive scalar problem.
\newblock {\em Physical Review E}, 60(4):4164, 1999.

\bibitem{voth2003mixing}
Greg~A Voth, TC~Saint, Greg Dobler, and Jerry~P Gollub.
\newblock Mixing rates and symmetry breaking in two-dimensional chaotic flow.
\newblock {\em Physics of Fluids}, 15(9):2560--2566, 2003.

\bibitem{rothstein1999persistent}
D~Rothstein, E~Henry, and Jerry~P Gollub.
\newblock Persistent patterns in transient chaotic fluid mixing.
\newblock {\em Nature}, 401(6755):770--772, 1999.

\bibitem{ran2021bacteria}
Ranjiangshang Ran, Quentin Brosseau, Brendan~C Blackwell, Boyang Qin, Rebecca~L
  Winter, and Paulo~E Arratia.
\newblock Bacteria hinder large-scale transport and enhance small-scale mixing
  in time-periodic flows.
\newblock {\em Proceedings of the National Academy of Sciences},
  118(40):e2108548118, 2021.

\bibitem{shraiman2000scalar}
Boris~I Shraiman and Eric~D Siggia.
\newblock Scalar turbulence.
\newblock {\em Nature}, 405(6787):639--646, 2000.

\bibitem{sreenivasan2019turbulent}
Katepalli~R Sreenivasan.
\newblock Turbulent mixing: A perspective.
\newblock {\em Proceedings of the National Academy of Sciences},
  116(37):18175--18183, 2019.

\bibitem{ottino1982description}
JM~Ottino.
\newblock Description of mixing with diffusion and reaction in terms of the
  concept of material surfaces.
\newblock {\em Journal of Fluid Mechanics}, 114:83--103, 1982.

\bibitem{shadden2006lagrangian}
Shawn~C Shadden, John~O Dabiri, and Jerrold~E Marsden.
\newblock Lagrangian analysis of fluid transport in empirical vortex ring
  flows.
\newblock {\em Physics of fluids}, 18(4):047105, 2006.

\bibitem{peng2009transport}
J~Peng and JO~Dabiri.
\newblock Transport of inertial particles by lagrangian coherent structures:
  application to predator--prey interaction in jellyfish feeding.
\newblock {\em Journal of Fluid Mechanics}, 623:75--84, 2009.

\bibitem{zurawik2016revisiting}
Tomasz~Micha{\l} {\.Z}urawik, Adam Pomorski, Agnieszka Belczyk-Ciesielska,
  Gra{\.z}yna Goch, Katarzyna Nied{\'z}wiedzka, R{\'o}{\.z}a Kucharczyk, Artur
  Kr{\k{e}}{\.z}el, and Wojciech Bal.
\newblock Revisiting mitochondrial ph with an improved algorithm for
  calibration of the ratiometric 5 (6)-carboxy-snarf-1 probe reveals
  anticooperative reaction with h+ ions and warrants further studies of
  organellar ph.
\newblock {\em PLoS One}, 11(8):e0161353, 2016.

\bibitem{baldyga1999turbulent}
Jerzy Ba{\l}dyga and John~R Bourne.
\newblock {\em Turbulent mixing and chemical reactions}.
\newblock John Wiley \& Sons, 1999.

\bibitem{song2003experimental}
Helen Song, Michelle~R Bringer, Joshua~D Tice, Cory~J Gerdts, and Rustem~F
  Ismagilov.
\newblock Experimental test of scaling of mixing by chaotic advection in
  droplets moving through microfluidic channels.
\newblock {\em Applied physics letters}, 83(22):4664--4666, 2003.

\bibitem{song2003millisecond}
Helen Song and Rustem~F Ismagilov.
\newblock Millisecond kinetics on a microfluidic chip using nanoliters of
  reagents.
\newblock {\em Journal of the American Chemical Society}, 125(47):14613--14619,
  2003.

\bibitem{teh2008droplet}
Shia-Yen Teh, Robert Lin, Lung-Hsin Hung, and Abraham~P Lee.
\newblock Droplet microfluidics.
\newblock {\em Lab on a Chip}, 8(2):198--220, 2008.

\bibitem{i2011droplet}
Xavier~Casadevall i~Solvas and Andrew DeMello.
\newblock Droplet microfluidics: recent developments and future applications.
\newblock {\em Chemical Communications}, 47(7):1936--1942, 2011.

\bibitem{guo2012droplet}
Mira~T Guo, Assaf Rotem, John~A Heyman, and David~A Weitz.
\newblock Droplet microfluidics for high-throughput biological assays.
\newblock {\em Lab on a Chip}, 12(12):2146--2155, 2012.

\bibitem{shang2017emerging}
Luoran Shang, Yao Cheng, and Yuanjin Zhao.
\newblock Emerging droplet microfluidics.
\newblock {\em Chemical reviews}, 117(12):7964--8040, 2017.

\bibitem{aquino2023fluid}
Tom\'as Aquino, Tanguy Le~Borgne, and Joris Heyman.
\newblock Fluid--solid reaction in porous media as a chaotic restart process.
\newblock {\em Phys. Rev. Lett.}, 130:264001, Jun 2023.

\bibitem{krummel2013visualizing}
Amber~T Krummel, Sujit~S Datta, Stefan M{\"u}nster, and David~A Weitz.
\newblock Visualizing multiphase flow and trapped fluid configurations in a
  model three-dimensional porous medium.
\newblock {\em AIChE Journal}, 59(3):1022--1029, 2013.

\bibitem{datta2013a}
Sujit~S Datta, Harry Chiang, TS~Ramakrishnan, and David~A Weitz.
\newblock Spatial fluctuations of fluid velocities in flow through a
  three-dimensional porous medium.
\newblock {\em {Physical Review Letters}}, 111(6):064501, 2013.

\bibitem{datta2013drainage}
Sujit~S Datta and David~A Weitz.
\newblock Drainage in a model stratified porous medium.
\newblock {\em EPL (Europhysics Letters)}, 101(1):14002, 2013.

\bibitem{shaqfeh1996}
Eric~SG Shaqfeh.
\newblock Purely elastic instabilities in viscometric flows.
\newblock {\em Annual Review of Fluid Mechanics}, 28:129--185, 1996.

\bibitem{galindo2012}
Francisco~J Galindo-Rosales, Laura Campo-Dea{\~n}o, FT~Pinho, E~Van~Bokhorst,
  PJ~Hamersma, M{\'o}nica~SN Oliveira, and MA~Alves.
\newblock Microfluidic systems for the analysis of viscoelastic fluid flow
  phenomena in porous media.
\newblock {\em Microfluidics and Nanofluidics}, 12(1-4):485--498, 2012.

\bibitem{clarke2016}
Andrew Clarke, Andrew~M Howe, Jonathan Mitchell, John Staniland, Laurence~A
  Hawkes, et~al.
\newblock How viscoelastic-polymer flooding enhances displacement efficiency.
\newblock {\em SPE Journal}, 21(03):675--687, 2016.

\bibitem{pan2013}
L~Pan, A~Morozov, C~Wagner, and PE~Arratia.
\newblock Nonlinear elastic instability in channel flows at low {R}eynolds
  numbers.
\newblock {\em {Physical Review Letters}}, 110(17):174502, 2013.

\bibitem{gendron2008diffusion}
P-O Gendron, F~Avaltroni, and KJ~Wilkinson.
\newblock Diffusion coefficients of several rhodamine derivatives as determined
  by pulsed field gradient--nuclear magnetic resonance and fluorescence
  correlation spectroscopy.
\newblock {\em Journal of fluorescence}, 18(6):1093--1101, 2008.

\bibitem{thielicke2014pivlab}
William Thielicke and Eize Stamhuis.
\newblock Pivlab--towards user-friendly, affordable and accurate digital
  particle image velocimetry in matlab.
\newblock {\em Journal of Open Research Software}, 2(1):e30, 2014.

\bibitem{song2003millisecondreaction}
Helen Song and Rustem~F Ismagilov.
\newblock Millisecond kinetics on a microfluidic chip using nanoliters of
  reagents.
\newblock {\em Journal of the American Chemical Society}, 125(47):14613--14619,
  2003.

\bibitem{groisman2004elastic}
A.~Groisman and V.~Steinberg.
\newblock {Elastic turbulence in curvilinear flows of polymer solutions}.
\newblock {\em New J. Phys.}, 6(1):29, 2004.

\bibitem{jha2020universal}
Narsing~K Jha and Victor Steinberg.
\newblock Universal coherent structures of elastic turbulence in straight
  channel with viscoelastic fluid flow.
\newblock {\em arXiv preprint arXiv:2009.12258}, 2020.

\bibitem{qin2017characterizing}
B.~Qin and P.~E. Arratia.
\newblock {Characterizing elastic turbulence in channel flows at low Reynolds
  number}.
\newblock {\em Phys. Rev. Fluids}, 2(8):083302, 2017.

\bibitem{vanBuel2018elastic}
R~van Buel, C~Schaaf, and H~Stark.
\newblock Elastic turbulence in two-dimensional taylor-couette flows.
\newblock {\em Europhys. Lett.}, 124(1):14001, 2018.

\bibitem{vanBuel2020active}
Reinier van Buel and Holger Stark.
\newblock Active open-loop control of elastic turbulence.
\newblock {\em Sci. Rep.}, 10(1):1--9, 2020.

\end{thebibliography}
\end{document}